%\magnification=\magstep1
\hsize=31pc 
\vsize=49pc 
\lineskip=0pt 
\parskip=0pt plus 1pt 
\hfuzz=1pt   
\vfuzz=2pt 
\pretolerance=2500 
\tolerance=5000 
\vbadness=5000 
\hbadness=5000 
\widowpenalty=500 
\clubpenalty=200 
\brokenpenalty=500 
\predisplaypenalty=200 
\voffset=-1pc 
\nopagenumbers      
\catcode`@=11 
\newif\ifams 
\amsfalse %\amstrue si ¤ suivant utilise
% 
%%%%%%%%%%%%%%%%%%%%%%%%%%%%%%%%%%%%%%%%%%%%%%%%%%%%%%%%%%%%% 
%                                                           % 
%  The following section may be commented out and           % 
%  \ifams set to either \amstrue to use the AMS fonts       % 
%  or \amsfalse if they are not available                   % 
%                                                           % 
%%%%%%%%%%%%%%%%%%%%%%%%%%%%%%%%%%%%%%%%%%%%%%%%%%%%%%%%%%%%% 
% 
%\def\Yesreply{Y } 
%\def\Noreply{N } 
%\def\yesreply{y } 
%\def\noreply{n } 
%\newif\ifnotyorn 
%\message{Do you want to use AMSfonts, msam and msbm? Y or N: }% 
%\loop 
%\read-1 to \reply 
%\ifx\reply\yesreply\global\amstrue\notyornfalse 
%\else\ifx\reply\Yesreply\global\amstrue\notyornfalse 
%\else\ifx\reply\noreply\global\amsfalse\notyornfalse 
%\else\ifx\reply\Noreply\global\amsfalse\notyornfalse 
%\else\notyorntrue 
%\message{Please type y or Y  (Yes) or n or N (No)}\fi\fi\fi\fi 
%\ifnotyorn\repeat 
%%%%%%%%%%%%%%%%%%%%%%%%%%%%%%%%%%%%%%%%%%%%%%%%%%%%%%%%%%%% 
% 
\newfam\bdifam 
\newfam\bsyfam 
\newfam\bssfam 
\newfam\msafam 
\newfam\msbfam 
\newif\ifxxpt    
\newif\ifxviipt  
\newif\ifxivpt   
\newif\ifxiipt   
\newif\ifxipt    
\newif\ifxpt     
\newif\ifixpt    
\newif\ifviiipt  
\newif\ifviipt   
\newif\ifvipt    
\newif\ifvpt     
% 
% Headings in 20pt, 17pt or 14pt 
% 
\def\headsize#1#2{\def\headb@seline{#2}% 
                \ifnum#1=20\def\HEAD{twenty}% 
                           \def\smHEAD{twelve}% 
                           \def\vsHEAD{nine}% 
                           \ifxxpt\else\xdef\f@ntsize{\HEAD}% 
                           \def\m@g{4}\def\s@ze{20.74}% 
                           \loadheadfonts\xxpttrue\fi 
                           \ifxiipt\else\xdef\f@ntsize{\smHEAD}% 
                           \def\m@g{1}\def\s@ze{12}% 
                           \loadxiiptfonts\xiipttrue\fi 
                           \ifixpt\else\xdef\f@ntsize{\vsHEAD}% 
                           \def\s@ze{9}% 
                           \loadsmallfonts\ixpttrue\fi 
                      \else 
                \ifnum#1=17\def\HEAD{seventeen}% 
                           \def\smHEAD{eleven}% 
                           \def\vsHEAD{eight}% 
                           \ifxviipt\else\xdef\f@ntsize{\HEAD}% 
                           \def\m@g{3}\def\s@ze{17.28}% 
                           \loadheadfonts\xviipttrue\fi 
                           \ifxipt\else\xdef\f@ntsize{\smHEAD}% 
                           \loadxiptfonts\xipttrue\fi 
                           \ifviiipt\else\xdef\f@ntsize{\vsHEAD}% 
                           \def\s@ze{8}% 
                           \loadsmallfonts\viiipttrue\fi 
                      \else\def\HEAD{fourteen}% 
                           \def\smHEAD{ten}% 
                           \def\vsHEAD{seven}% 
                           \ifxivpt\else\xdef\f@ntsize{\HEAD}% 
                           \def\m@g{2}\def\s@ze{14.4}% 
                           \loadheadfonts\xivpttrue\fi 
                           \ifxpt\else\xdef\f@ntsize{\smHEAD}% 
                           \def\s@ze{10}% 
                           \loadxptfonts\xpttrue\fi 
                           \ifviipt\else\xdef\f@ntsize{\vsHEAD}% 
                           \def\s@ze{7}% 
                           \loadviiptfonts\viipttrue\fi 
                \ifnum#1=14\else 
                \message{Header size should be 20, 17 or 14 point 
                              will now default to 14pt}\fi 
                \fi\fi\headfonts} 
% 
% Text in 12pt, 11pt or 10pt  
% 
\def\textsize#1#2{\def\textb@seline{#2}% 
                 \ifnum#1=12\def\TEXT{twelve}% 
                           \def\smTEXT{eight}% 
                           \def\vsTEXT{six}% 
                           \ifxiipt\else\xdef\f@ntsize{\TEXT}% 
                           \def\m@g{1}\def\s@ze{12}% 
                           \loadxiiptfonts\xiipttrue\fi 
                           \ifviiipt\else\xdef\f@ntsize{\smTEXT}% 
                           \def\s@ze{8}% 
                           \loadsmallfonts\viiipttrue\fi 
                           \ifvipt\else\xdef\f@ntsize{\vsTEXT}% 
                           \def\s@ze{6}% 
                           \loadviptfonts\vipttrue\fi 
                      \else 
                \ifnum#1=11\def\TEXT{eleven}% 
                           \def\smTEXT{seven}% 
                           \def\vsTEXT{five}% 
                           \ifxipt\else\xdef\f@ntsize{\TEXT}% 
                           \def\s@ze{11}% 
                           \loadxiptfonts\xipttrue\fi 
                           \ifviipt\else\xdef\f@ntsize{\smTEXT}% 
                           \loadviiptfonts\viipttrue\fi 
                           \ifvpt\else\xdef\f@ntsize{\vsTEXT}% 
                           \def\s@ze{5}% 
                           \loadvptfonts\vpttrue\fi 
                      \else\def\TEXT{ten}% 
                           \def\smTEXT{seven}% 
                           \def\vsTEXT{five}% 
                           \ifxpt\else\xdef\f@ntsize{\TEXT}% 
                           \loadxptfonts\xpttrue\fi 
                           \ifviipt\else\xdef\f@ntsize{\smTEXT}% 
                           \def\s@ze{7}% 
                           \loadviiptfonts\viipttrue\fi 
                           \ifvpt\else\xdef\f@ntsize{\vsTEXT}% 
                           \def\s@ze{5}% 
                           \loadvptfonts\vpttrue\fi 
                \ifnum#1=10\else 
                \message{Text size should be 12, 11 or 10 point 
                              will now default to 10pt}\fi 
                \fi\fi\textfonts} 
% 
% Small sized material in 10pt, 9pt or 8pt 
% 
\def\smallsize#1#2{\def\smallb@seline{#2}% 
                 \ifnum#1=10\def\SMALL{ten}% 
                           \def\smSMALL{seven}% 
                           \def\vsSMALL{five}% 
                           \ifxpt\else\xdef\f@ntsize{\SMALL}% 
                           \loadxptfonts\xpttrue\fi 
                           \ifviipt\else\xdef\f@ntsize{\smSMALL}% 
                           \def\s@ze{7}% 
                           \loadviiptfonts\viipttrue\fi 
                           \ifvpt\else\xdef\f@ntsize{\vsSMALL}% 
                           \def\s@ze{5}% 
                           \loadvptfonts\vpttrue\fi 
                       \else 
                 \ifnum#1=9\def\SMALL{nine}% 
                           \def\smSMALL{six}% 
                           \def\vsSMALL{five}% 
                           \ifixpt\else\xdef\f@ntsize{\SMALL}% 
                           \def\s@ze{9}% 
                           \loadsmallfonts\ixpttrue\fi 
                           \ifvipt\else\xdef\f@ntsize{\smSMALL}% 
                           \def\s@ze{6}% 
                           \loadviptfonts\vipttrue\fi 
                           \ifvpt\else\xdef\f@ntsize{\vsSMALL}% 
                           \def\s@ze{5}% 
                           \loadvptfonts\vpttrue\fi 
                       \else 
                           \def\SMALL{eight}% 
                           \def\smSMALL{six}% 
                           \def\vsSMALL{five}% 
                           \ifviiipt\else\xdef\f@ntsize{\SMALL}% 
                           \def\s@ze{8}% 
                           \loadsmallfonts\viiipttrue\fi 
                           \ifvipt\else\xdef\f@ntsize{\smSMALL}% 
                           \def\s@ze{6}% 
                           \loadviptfonts\vipttrue\fi 
                           \ifvpt\else\xdef\f@ntsize{\vsSMALL}% 
                           \def\s@ze{5}% 
                           \loadvptfonts\vpttrue\fi 
                 \ifnum#1=8\else\message{Small size should be 10, 9 or  
                            8 point will now default to 8pt}\fi 
                \fi\fi\smallfonts} 
\def\F@nt{\expandafter\font\csname} 
\def\Sk@w{\expandafter\skewchar\csname} 
\def\@nd{\endcsname} 
\def\@step#1{ scaled \magstep#1} 
\def\@half{ scaled \magstephalf} 
\def\@t#1{ at #1pt} 
% 
% For 14, 17 and 20 point fonts use \loadheadfonts 
% 
\def\loadheadfonts{\bigf@nts 
\F@nt \f@ntsize bdi\@nd=cmmib10 \@t{\s@ze}% 
\Sk@w \f@ntsize bdi\@nd='177 
\F@nt \f@ntsize bsy\@nd=cmbsy10 \@t{\s@ze}% 
\Sk@w \f@ntsize bsy\@nd='60 
\F@nt \f@ntsize bss\@nd=cmssbx10 \@t{\s@ze}} 
% 
% For 12 point fonts use \loadxiiptfonts 
% 
\def\loadxiiptfonts{\bigf@nts 
\F@nt \f@ntsize bdi\@nd=cmmib10 \@step{\m@g}% 
\Sk@w \f@ntsize bdi\@nd='177 
\F@nt \f@ntsize bsy\@nd=cmbsy10 \@step{\m@g}% 
\Sk@w \f@ntsize bsy\@nd='60 
\F@nt \f@ntsize bss\@nd=cmssbx10 \@step{\m@g}} 
% 
% For 11 point fonts use \loadxiptfonts 
% 
\def\loadxiptfonts{% 
\font\elevenrm=cmr10 \@half 
\font\eleveni=cmmi10 \@half 
\skewchar\eleveni='177 
\font\elevensy=cmsy10 \@half 
\skewchar\elevensy='60 
\font\elevenex=cmex10 \@half 
\font\elevenit=cmti10 \@half 
\font\elevensl=cmsl10 \@half 
\font\elevenbf=cmbx10 \@half 
\font\eleventt=cmtt10 \@half 
\ifams\font\elevenmsa=msam10 \@half 
\font\elevenmsb=msbm10 \@half\else\fi 
\font\elevenbdi=cmmib10 \@half 
\skewchar\elevenbdi='177 
\font\elevenbsy=cmbsy10 \@half 
\skewchar\elevenbsy='60 
\font\elevenbss=cmssbx10 \@half} 
% 
% For 10 point fonts use \loadxptfonts 
% 
\def\loadxptfonts{% 
\font\tenbdi=cmmib10 
\skewchar\tenbdi='177 
\font\tenbsy=cmbsy10  
\skewchar\tenbsy='60 
\ifams\font\tenmsa=msam10  
\font\tenmsb=msbm10\else\fi 
\font\tenbss=cmssbx10}%  
% 
% For 8 and 9 point fonts use \loadsmallfonts 
% 
\def\loadsmallfonts{\smallf@nts 
\ifams 
\F@nt \f@ntsize ex\@nd=cmex\s@ze 
\else 
\F@nt \f@ntsize ex\@nd=cmex10\fi 
\F@nt \f@ntsize it\@nd=cmti\s@ze 
\F@nt \f@ntsize sl\@nd=cmsl\s@ze 
\F@nt \f@ntsize tt\@nd=cmtt\s@ze} 
% 
% For 7 point fonts use \loadviiptfonts 
% 
\def\loadviiptfonts{% 
\font\sevenit=cmti7 
\font\sevensl=cmsl8 at 7pt 
\ifams\font\sevenmsa=msam7  
\font\sevenmsb=msbm7 
\font\sevenex=cmex7 
\font\sevenbsy=cmbsy7 
\font\sevenbdi=cmmib7\else 
\font\sevenex=cmex10 
\font\sevenbsy=cmbsy10 at 7pt 
\font\sevenbdi=cmmib10 at 7pt\fi 
\skewchar\sevenbsy='60 
\skewchar\sevenbdi='177 
\font\sevenbss=cmssbx10 at 7pt}%  
% 
%  For 6 point fonts use \loadviptfonts 
% 
\def\loadviptfonts{\smallf@nts 
\ifams\font\sixex=cmex7 at 6pt\else 
\font\sixex=cmex10\fi 
\font\sixit=cmti7 at 6pt} 
% 
% For 5 point fonts use \loadvptfonts 
% 
\def\loadvptfonts{% 
\font\fiveit=cmti7 at 5pt 
\ifams\font\fiveex=cmex7 at 5pt 
\font\fivebdi=cmmib5 
\font\fivebsy=cmbsy5 
\font\fivemsa=msam5  
\font\fivemsb=msbm5\else 
\font\fiveex=cmex10 
\font\fivebdi=cmmib10 at 5pt 
\font\fivebsy=cmbsy10 at 5pt\fi 
\skewchar\fivebdi='177 
\skewchar\fivebsy='60 
\font\fivebss=cmssbx10 at 5pt} 
\def\bigf@nts{% 
\F@nt \f@ntsize rm\@nd=cmr10 \@step{\m@g}% 
\F@nt \f@ntsize i\@nd=cmmi10 \@step{\m@g}% 
\Sk@w \f@ntsize i\@nd='177 
\F@nt \f@ntsize sy\@nd=cmsy10 \@step{\m@g}% 
\Sk@w \f@ntsize sy\@nd='60 
\F@nt \f@ntsize ex\@nd=cmex10 \@step{\m@g}% 
\F@nt \f@ntsize it\@nd=cmti10 \@step{\m@g}% 
\F@nt \f@ntsize sl\@nd=cmsl10 \@step{\m@g}% 
\F@nt \f@ntsize bf\@nd=cmbx10 \@step{\m@g}% 
\F@nt \f@ntsize tt\@nd=cmtt10 \@step{\m@g}% 
\ifams 
\F@nt \f@ntsize msa\@nd=msam10 \@step{\m@g}% 
\F@nt \f@ntsize msb\@nd=msbm10 \@step{\m@g}\else\fi} 
\def\smallf@nts{% 
\F@nt \f@ntsize rm\@nd=cmr\s@ze 
\F@nt \f@ntsize i\@nd=cmmi\s@ze  
\Sk@w \f@ntsize i\@nd='177 
\F@nt \f@ntsize sy\@nd=cmsy\s@ze 
\Sk@w \f@ntsize sy\@nd='60 
\F@nt \f@ntsize bf\@nd=cmbx\s@ze  
\ifams 
\F@nt \f@ntsize bdi\@nd=cmmib\s@ze  
\F@nt \f@ntsize bsy\@nd=cmbsy\s@ze  
\F@nt \f@ntsize msa\@nd=msam\s@ze  
\F@nt \f@ntsize msb\@nd=msbm\s@ze 
\else 
\F@nt \f@ntsize bdi\@nd=cmmib10 \@t{\s@ze}%  
\F@nt \f@ntsize bsy\@nd=cmbsy10 \@t{\s@ze}\fi  
\Sk@w \f@ntsize bdi\@nd='177 
\Sk@w \f@ntsize bsy\@nd='60 
\F@nt \f@ntsize bss\@nd=cmssbx10 \@t{\s@ze}}%  
% 
% Fonts for headings  
% 
\def\headfonts{% 
\textfont0=\csname\HEAD rm\@nd         
\scriptfont0=\csname\smHEAD rm\@nd 
\scriptscriptfont0=\csname\vsHEAD rm\@nd 
\def\rm{\fam0\csname\HEAD rm\@nd 
\def\sc{\csname\smHEAD rm\@nd}}% 
\textfont1=\csname\HEAD i\@nd          
\scriptfont1=\csname\smHEAD i\@nd 
\scriptscriptfont1=\csname\vsHEAD i\@nd 
\textfont2=\csname\HEAD sy\@nd         
\scriptfont2=\csname\smHEAD sy\@nd 
\scriptscriptfont2=\csname\vsHEAD sy\@nd 
\textfont3=\csname\HEAD ex\@nd         
\scriptfont3=\csname\smHEAD ex\@nd 
\scriptscriptfont3=\csname\smHEAD ex\@nd 
\textfont\itfam=\csname\HEAD it\@nd    
\scriptfont\itfam=\csname\smHEAD it\@nd 
\scriptscriptfont\itfam=\csname\vsHEAD it\@nd 
\def\it{\fam\itfam\csname\HEAD it\@nd 
\def\sc{\csname\smHEAD it\@nd}}% 
\textfont\slfam=\csname\HEAD sl\@nd    
\def\sl{\fam\slfam\csname\HEAD sl\@nd 
\def\sc{\csname\smHEAD sl\@nd}}% 
\textfont\bffam=\csname\HEAD bf\@nd    
\scriptfont\bffam=\csname\smHEAD bf\@nd 
\scriptscriptfont\bffam=\csname\vsHEAD bf\@nd 
\def\bf{\fam\bffam\csname\HEAD bf\@nd 
\def\sc{\csname\smHEAD bf\@nd}}% 
\textfont\ttfam=\csname\HEAD tt\@nd    
\def\tt{\fam\ttfam\csname\HEAD tt\@nd}% 
\textfont\bdifam=\csname\HEAD bdi\@nd  
\scriptfont\bdifam=\csname\smHEAD bdi\@nd 
\scriptscriptfont\bdifam=\csname\vsHEAD bdi\@nd 
\def\bdi{\fam\bdifam\csname\HEAD bdi\@nd}% 
\textfont\bsyfam=\csname\HEAD bsy\@nd  
\scriptfont\bsyfam=\csname\smHEAD bsy\@nd 
\def\bsy{\fam\bsyfam\csname\HEAD bsy\@nd}% 
\textfont\bssfam=\csname\HEAD bss\@nd  
\scriptfont\bssfam=\csname\smHEAD bss\@nd 
\scriptscriptfont\bssfam=\csname\vsHEAD bss\@nd 
\def\bss{\fam\bssfam\csname\HEAD bss\@nd}% 
\ifams 
\textfont\msafam=\csname\HEAD msa\@nd  
\scriptfont\msafam=\csname\smHEAD msa\@nd 
\scriptscriptfont\msafam=\csname\vsHEAD msa\@nd 
\textfont\msbfam=\csname\HEAD msb\@nd  
\scriptfont\msbfam=\csname\smHEAD msb\@nd 
\scriptscriptfont\msbfam=\csname\vsHEAD msb\@nd 
\else\fi 
\normalbaselineskip=\headb@seline pt% 
\setbox\strutbox=\hbox{\vrule height.7\normalbaselineskip  
depth.3\baselineskip width0pt}% 
\def\sc{\csname\smHEAD rm\@nd}\normalbaselines\bf} 
% 
% Fonts for text 
% 
\def\textfonts{% 
\textfont0=\csname\TEXT rm\@nd         
\scriptfont0=\csname\smTEXT rm\@nd 
\scriptscriptfont0=\csname\vsTEXT rm\@nd 
\def\rm{\fam0\csname\TEXT rm\@nd 
\def\sc{\csname\smTEXT rm\@nd}}% 
\textfont1=\csname\TEXT i\@nd          
\scriptfont1=\csname\smTEXT i\@nd 
\scriptscriptfont1=\csname\vsTEXT i\@nd 
\textfont2=\csname\TEXT sy\@nd         
\scriptfont2=\csname\smTEXT sy\@nd 
\scriptscriptfont2=\csname\vsTEXT sy\@nd 
\textfont3=\csname\TEXT ex\@nd         
\scriptfont3=\csname\smTEXT ex\@nd 
\scriptscriptfont3=\csname\smTEXT ex\@nd 
\textfont\itfam=\csname\TEXT it\@nd    
\scriptfont\itfam=\csname\smTEXT it\@nd 
\scriptscriptfont\itfam=\csname\vsTEXT it\@nd 
\def\it{\fam\itfam\csname\TEXT it\@nd 
\def\sc{\csname\smTEXT it\@nd}}% 
\textfont\slfam=\csname\TEXT sl\@nd    
\def\sl{\fam\slfam\csname\TEXT sl\@nd 
\def\sc{\csname\smTEXT sl\@nd}}% 
\textfont\bffam=\csname\TEXT bf\@nd    
\scriptfont\bffam=\csname\smTEXT bf\@nd 
\scriptscriptfont\bffam=\csname\vsTEXT bf\@nd 
\def\bf{\fam\bffam\csname\TEXT bf\@nd 
\def\sc{\csname\smTEXT bf\@nd}}% 
\textfont\ttfam=\csname\TEXT tt\@nd    
\def\tt{\fam\ttfam\csname\TEXT tt\@nd}% 
\textfont\bdifam=\csname\TEXT bdi\@nd  
\scriptfont\bdifam=\csname\smTEXT bdi\@nd 
\scriptscriptfont\bdifam=\csname\vsTEXT bdi\@nd 
\def\bdi{\fam\bdifam\csname\TEXT bdi\@nd}% 
\textfont\bsyfam=\csname\TEXT bsy\@nd  
\scriptfont\bsyfam=\csname\smTEXT bsy\@nd 
\def\bsy{\fam\bsyfam\csname\TEXT bsy\@nd}% 
\textfont\bssfam=\csname\TEXT bss\@nd  
\scriptfont\bssfam=\csname\smTEXT bss\@nd 
\scriptscriptfont\bssfam=\csname\vsTEXT bss\@nd 
\def\bss{\fam\bssfam\csname\TEXT bss\@nd}% 
\ifams 
\textfont\msafam=\csname\TEXT msa\@nd  
\scriptfont\msafam=\csname\smTEXT msa\@nd 
\scriptscriptfont\msafam=\csname\vsTEXT msa\@nd 
\textfont\msbfam=\csname\TEXT msb\@nd  
\scriptfont\msbfam=\csname\smTEXT msb\@nd 
\scriptscriptfont\msbfam=\csname\vsTEXT msb\@nd 
\else\fi 
\normalbaselineskip=\textb@seline pt 
\setbox\strutbox=\hbox{\vrule height.7\normalbaselineskip  
depth.3\baselineskip width0pt}% 
\everymath{}% 
\def\sc{\csname\smTEXT rm\@nd}\normalbaselines\rm} 
% 
% Fonts for small material (captions, footnotes etc) 
% 
\def\smallfonts{% 
\textfont0=\csname\SMALL rm\@nd         
\scriptfont0=\csname\smSMALL rm\@nd 
\scriptscriptfont0=\csname\vsSMALL rm\@nd 
\def\rm{\fam0\csname\SMALL rm\@nd 
\def\sc{\csname\smSMALL rm\@nd}}% 
\textfont1=\csname\SMALL i\@nd          
\scriptfont1=\csname\smSMALL i\@nd 
\scriptscriptfont1=\csname\vsSMALL i\@nd 
\textfont2=\csname\SMALL sy\@nd         
\scriptfont2=\csname\smSMALL sy\@nd 
\scriptscriptfont2=\csname\vsSMALL sy\@nd 
\textfont3=\csname\SMALL ex\@nd         
\scriptfont3=\csname\smSMALL ex\@nd 
\scriptscriptfont3=\csname\smSMALL ex\@nd 
\textfont\itfam=\csname\SMALL it\@nd    
\scriptfont\itfam=\csname\smSMALL it\@nd 
\scriptscriptfont\itfam=\csname\vsSMALL it\@nd 
\def\it{\fam\itfam\csname\SMALL it\@nd 
\def\sc{\csname\smSMALL it\@nd}}% 
\textfont\slfam=\csname\SMALL sl\@nd    
\def\sl{\fam\slfam\csname\SMALL sl\@nd 
\def\sc{\csname\smSMALL sl\@nd}}% 
\textfont\bffam=\csname\SMALL bf\@nd    
\scriptfont\bffam=\csname\smSMALL bf\@nd 
\scriptscriptfont\bffam=\csname\vsSMALL bf\@nd 
\def\bf{\fam\bffam\csname\SMALL bf\@nd 
\def\sc{\csname\smSMALL bf\@nd}}% 
\textfont\ttfam=\csname\SMALL tt\@nd    
\def\tt{\fam\ttfam\csname\SMALL tt\@nd}% 
\textfont\bdifam=\csname\SMALL bdi\@nd  
\scriptfont\bdifam=\csname\smSMALL bdi\@nd 
\scriptscriptfont\bdifam=\csname\vsSMALL bdi\@nd 
\def\bdi{\fam\bdifam\csname\SMALL bdi\@nd}% 
\textfont\bsyfam=\csname\SMALL bsy\@nd  
\scriptfont\bsyfam=\csname\smSMALL bsy\@nd 
\def\bsy{\fam\bsyfam\csname\SMALL bsy\@nd}% 
\textfont\bssfam=\csname\SMALL bss\@nd  
\scriptfont\bssfam=\csname\smSMALL bss\@nd 
\scriptscriptfont\bssfam=\csname\vsSMALL bss\@nd 
\def\bss{\fam\bssfam\csname\SMALL bss\@nd}% 
\ifams 
\textfont\msafam=\csname\SMALL msa\@nd  
\scriptfont\msafam=\csname\smSMALL msa\@nd 
\scriptscriptfont\msafam=\csname\vsSMALL msa\@nd 
\textfont\msbfam=\csname\SMALL msb\@nd  
\scriptfont\msbfam=\csname\smSMALL msb\@nd 
\scriptscriptfont\msbfam=\csname\vsSMALL msb\@nd 
\else\fi 
\normalbaselineskip=\smallb@seline pt% 
\setbox\strutbox=\hbox{\vrule height.7\normalbaselineskip  
depth.3\baselineskip width0pt}% 
\everymath{}% 
\def\sc{\csname\smSMALL rm\@nd}\normalbaselines\rm}% 
\everydisplay{\indenteddisplay 
   \gdef\labeltype{\eqlabel}}% 
% 
%%%%%%%%%%%%%%%%%%%%%%%%%%%%%%%%%%%%%%%%%%%%%%%%%%%%%%%%%%% 
%                                                         % 
%  Macros to define extra maths symbols                   % 
%                                                         % 
%%%%%%%%%%%%%%%%%%%%%%%%%%%%%%%%%%%%%%%%%%%%%%%%%%%%%%%%%%% 
% 
\def\hexnumber@#1{\ifcase#1 0\or 1\or 2\or 3\or 4\or 5\or 6\or 7\or 8\or 
 9\or A\or B\or C\or D\or E\or F\fi} 
\edef\bffam@{\hexnumber@\bffam} 
\edef\bdifam@{\hexnumber@\bdifam} 
\edef\bsyfam@{\hexnumber@\bsyfam} 
\def\undefine#1{\let#1\undefined} 
\def\newsymbol#1#2#3#4#5{\let\next@\relax 
 \ifnum#2=\thr@@\let\next@\bdifam@\else 
 \ifams 
 \ifnum#2=\@ne\let\next@\msafam@\else 
 \ifnum#2=\tw@\let\next@\msbfam@\fi\fi 
 \fi\fi 
 \mathchardef#1="#3\next@#4#5} 
\def\mathhexbox@#1#2#3{\relax 
 \ifmmode\mathpalette{}{\m@th\mathchar"#1#2#3}% 
 \else\leavevmode\hbox{$\m@th\mathchar"#1#2#3$}\fi} 

\def\bi#1{{\fam\bdifam\relax#1}} 
% 
% If file amsmacro is not in current directory 
% or somewhere with set path add path before 
% file name in following line 
% 
\ifams\input amsmacro\fi 
% 
% Bold italic Greek characters 
% 
\newsymbol\bitGamma 3000 
\newsymbol\bitDelta 3001 
\newsymbol\bitTheta 3002 
\newsymbol\bitLambda 3003 
\newsymbol\bitXi 3004 
\newsymbol\bitPi 3005 
\newsymbol\bitSigma 3006 
\newsymbol\bitUpsilon 3007 
\newsymbol\bitPhi 3008 
\newsymbol\bitPsi 3009 
\newsymbol\bitOmega 300A 
\newsymbol\balpha 300B 
\newsymbol\bbeta 300C 
\newsymbol\bgamma 300D 
\newsymbol\bdelta 300E 
\newsymbol\bepsilon 300F 
\newsymbol\bzeta 3010 
\newsymbol\bfeta 3011 
\newsymbol\btheta 3012 
\newsymbol\biota 3013 
\newsymbol\bkappa 3014 
\newsymbol\blambda 3015 
\newsymbol\bmu 3016 
\newsymbol\bnu 3017 
\newsymbol\bxi 3018 
\newsymbol\bpi 3019 
\newsymbol\brho 301A 
\newsymbol\bsigma 301B 
\newsymbol\btau 301C 
\newsymbol\bupsilon 301D 
\newsymbol\bphi 301E 
\newsymbol\bchi 301F 
\newsymbol\bpsi 3020 
\newsymbol\bomega 3021 
\newsymbol\bvarepsilon 3022 
\newsymbol\bvartheta 3023 
\newsymbol\bvaromega 3024 
\newsymbol\bvarrho 3025 
\newsymbol\bvarzeta 3026 
\newsymbol\bvarphi 3027 
\newsymbol\bpartial 3040 
\newsymbol\bell 3060 
\newsymbol\bimath 307B 
\newsymbol\bjmath 307C 
\mathchardef\binfty "0\bsyfam@31 
\mathchardef\bnabla "0\bsyfam@72 
\mathchardef\bdot "2\bsyfam@01 
\mathchardef\bGamma "0\bffam@00 
\mathchardef\bDelta "0\bffam@01 
\mathchardef\bTheta "0\bffam@02 
\mathchardef\bLambda "0\bffam@03 
\mathchardef\bXi "0\bffam@04 
\mathchardef\bPi "0\bffam@05 
\mathchardef\bSigma "0\bffam@06 
\mathchardef\bUpsilon "0\bffam@07 
\mathchardef\bPhi "0\bffam@08 
\mathchardef\bPsi "0\bffam@09 
\mathchardef\bOmega "0\bffam@0A 
\mathchardef\itGamma "0100 
\mathchardef\itDelta "0101 
\mathchardef\itTheta "0102 
\mathchardef\itLambda "0103 
\mathchardef\itXi "0104 
\mathchardef\itPi "0105 
\mathchardef\itSigma "0106 
\mathchardef\itUpsilon "0107 
\mathchardef\itPhi "0108 
\mathchardef\itPsi "0109 
\mathchardef\itOmega "010A 
\mathchardef\Gamma "0000 
\mathchardef\Delta "0001 
\mathchardef\Theta "0002 
\mathchardef\Lambda "0003 
\mathchardef\Xi "0004 
\mathchardef\Pi "0005 
\mathchardef\Sigma "0006 
\mathchardef\Upsilon "0007 
\mathchardef\Phi "0008 
\mathchardef\Psi "0009 
\mathchardef\Omega "000A 
% 
% Counter definitions 
% 
\newcount\firstpage  \firstpage=1  % start page no 
\newcount\jnl                      % journal no 
\newcount\secno                    % section number 
\newcount\subno                    % number of subsection 
\newcount\subsubno                 % number of subsubsection 
\newcount\appno                    % appendix number 
\newcount\tabno                    % table number 
\newcount\figno                    % figure number 
\newcount\countno                  % equation numbers 
\newcount\refno                    % reference number 
\newcount\eqlett     \eqlett=97    % equation letter 
\newif\ifletter 
\newif\ifwide 
\newif\ifnotfull 
\newif\ifaligned 
\newif\ifnumbysec   
\newif\ifappendix 
\newif\ifnumapp 
\newif\ifssf 
\newif\ifppt 
\newdimen\t@bwidth 
\newdimen\c@pwidth 
\newdimen\digitwidth                    %character width 
\newdimen\argwidth                      %argument width 
\newdimen\secindent    \secindent=5pc   %indentation of maths  
\newdimen\textind    \textind=16pt      %indentation of text 
\newdimen\tempval                       %temporary value 
\newskip\beforesecskip 
\def\beforesecspace{\vskip\beforesecskip\relax} 
\newskip\beforesubskip 
\def\beforesubspace{\vskip\beforesubskip\relax} 
\newskip\beforesubsubskip 
\def\beforesubsubspace{\vskip\beforesubsubskip\relax} 
\newskip\secskip 
\def\secspace{\vskip\secskip\relax} 
\newskip\subskip 
\def\subspace{\vskip\subskip\relax} 
\newskip\insertskip 
\def\insertspace{\vskip\insertskip\relax} 
\def\sp@ce{\ifx\next*\let\next=\@ssf 
               \else\let\next=\@nossf\fi\next} 
\def\@ssf#1{\nobreak\secspace\global\ssftrue\nobreak} 
\def\@nossf{\nobreak\secspace\nobreak\noindent\ignorespaces} 
\def\subsp@ce{\ifx\next*\let\next=\@sssf 
               \else\let\next=\@nosssf\fi\next} 
\def\@sssf#1{\nobreak\subspace\global\ssftrue\nobreak} 
\def\@nosssf{\nobreak\subspace\nobreak\noindent\ignorespaces} 
\beforesecskip=24pt plus12pt minus8pt 
\beforesubskip=12pt plus6pt minus4pt 
\beforesubsubskip=12pt plus6pt minus4pt 
\secskip=12pt plus 2pt minus 2pt 
\subskip=6pt plus3pt minus2pt 
\insertskip=18pt plus6pt minus6pt% 
\fontdimen16\tensy=2.7pt 
\fontdimen17\tensy=2.7pt 
% 
% Labels etc for cross referencing macros 
% 
\def\eqlabel{(\ifappendix\applett 
               \ifnumbysec\ifnum\secno>0 \the\secno\fi.\fi 
               \else\ifnumbysec\the\secno.\fi\fi\the\countno)} 
\def\seclabel{\ifappendix\ifnumapp\else\applett\fi 
    \ifnum\secno>0 \the\secno 
    \ifnumbysec\ifnum\subno>0.\the\subno\fi\fi\fi 
    \else\the\secno\fi\ifnum\subno>0.\the\subno 
         \ifnum\subsubno>0.\the\subsubno\fi\fi} 
\def\tablabel{\ifappendix\applett\fi\the\tabno} 
\def\figlabel{\ifappendix\applett\fi\the\figno} 
\def\gac{\global\advance\countno by 1} 
% 
% Redefinition of footnote macros to lose rule and remove indentation 
% 
 
\def\vfootnote#1{\insert\footins\bgroup 
\interlinepenalty=\interfootnotelinepenalty 
\splittopskip=\ht\strutbox % top baseline for broken footnotes 
\splitmaxdepth=\dp\strutbox \floatingpenalty=20000 
\leftskip=0pt \rightskip=0pt \spaceskip=0pt \xspaceskip=0pt% 
\noindent\smallfonts\rm #1\ \ignorespaces\footstrut\futurelet\next\fo@t} 
% 
% Redefinition of endinsert to give more controllable 
% space around  tables and figures 
% 
\def\endinsert{\egroup 
    \if@mid \dimen@=\ht0 \advance\dimen@ by\dp0 
       \advance\dimen@ by12\p@ \advance\dimen@ by\pagetotal 
       \ifdim\dimen@>\pagegoal \@midfalse\p@gefalse\fi\fi 
    \if@mid \insertspace \box0 \par \ifdim\lastskip<\insertskip 
    \removelastskip \penalty-200 \insertspace \fi 
    \else\insert\topins{\penalty100 
       \splittopskip=0pt \splitmaxdepth=\maxdimen  
       \floatingpenalty=0 
       \ifp@ge \dimen@=\dp0 
       \vbox to\vsize{\unvbox0 \kern-\dimen@}% 
       \else\box0\nobreak\insertspace\fi}\fi\endgroup}    
% 
% special macros for display equations 
% 
% for indentation of turned over lines in mathematics 
% 
\def\ind{\hbox to \secindent{\hfill}} 
% 
% for turned over equals sign to left of maths indent 
% 

% 
% for other signs to left of maths indent 
% 
 
% 
% displayed equation indented  
% 
\def\indeqn#1{\alignedfalse\displ@y\halign{\hbox to \displaywidth 
    {$\ind\@lign\displaystyle##\hfil$}\crcr #1\crcr}} 
% 
% displayed equation indented with alignments 
% 
\def\indalign#1{\alignedtrue\displ@y \tabskip=0pt  
  \halign to\displaywidth{\ind$\@lign\displaystyle{##}$\tabskip=0pt 
    &$\@lign\displaystyle{{}##}$\hfill\tabskip=\centering 
    &\llap{$\@lign\hbox{\rm##}$}\tabskip=0pt\crcr 
    #1\crcr}} 
\def\fl{{\hskip-\secindent}} 
\def\indenteddisplay#1$${\indispl@y{#1 }} 
\def\indispl@y#1{\disptest#1\eqalignno\eqalignno\disptest} 
\def\disptest#1\eqalignno#2\eqalignno#3\disptest{% 
    \ifx#3\eqalignno 
    \indalign#2% 
    \else\indeqn{#1}\fi$$} 
% 
% Roman small caps (if in Roman \sc gives small caps) 
% 
 
% 
% Italic small caps (if in italic \sc gives italic small caps) 
% 
 
% 
% Bold small caps (if in bold \sc gives bold small caps) 
% 
 
% 
% Small caps in maths 
% 
 
% 
% Miscellaneous definitions 
% 

\def\ns{\noalign{\vskip-3pt}}

% 
 
% 
% Bold h bar 
% 
\def\bhbar{\rlap{\kern1pt\raise.4ex\hbox{\bf\char'40}}\bi{h}} 
\def\case#1#2{{\textstyle{#1\over#2}}}

\def\d{{\rm d}}

\def\frac#1#2{{#1\over#2}} 
\ifams 
\def\lap{\lesssim} 
\def\gap{\gtrsim}

\let\leq=\leqslant 
\let\ge=\geqslant

\else

\def\gap{\;\lower3pt\hbox{$\buildrel > \over \sim$}\;}% 
\def\lap{\;\lower3pt\hbox{$\buildrel < \over \sim$}\;}\fi 
\def\i{{\rm i}} 
\chardef\ii="10 
\def\tqs{\hbox to 25pt{\hfil}}

\def\Or{\mathop{\rm O}\nolimits}

\def\Bbbone{1\kern-.22em {\rm l}} 
% 
% Primes to display summations and products  
% which also have sub or superscripts 
% 
\def\rp{\raise8pt\hbox{$\scriptstyle\prime$}} 
% 
% then use \sum^{...}_{...}\rp or \prod^{...}_{...}\rp. 
% 
% Shadow brackets 
% 
% Single brackets for normal size only 
% 

% 
% Variable size for display style 
% 
\def\[#1\]{\setbox0=\hbox{$\dsty#1$}\argwidth=\wd0 
    \setbox0=\hbox{$\left[\box0\right]$}\advance\argwidth by -\wd0 
    \left[\kern.3\argwidth\box0\kern.3\argwidth\right]} 
% 
% Variable size for text style 
% 
\def\lsb#1\rsb{\setbox0=\hbox{$#1$}\argwidth=\wd0 
    \setbox0=\hbox{$\left[\box0\right]$}\advance\argwidth by -\wd0 
    \left[\kern.3\argwidth\box0\kern.3\argwidth\right]} 
% 
 
% 
% Square for end of theorems 
% 
 
% 
\def\pt(#1){({\it #1\/})} 
\let\dsty=\displaystyle

% 
% Definition for Nuclear Physics Keyword abstract 
% 
\def\reactions#1{\vskip 12pt plus2pt minus2pt%              
\vbox{\hbox{\kern\secindent\vrule\kern12pt% 
\vbox{\kern0.5pt\vbox{\hsize=24pc\parindent=0pt\smallfonts\rm NUCLEAR  
REACTIONS\strut\quad #1\strut}\kern0.5pt}\kern12pt\vrule}}} 
% 
% Definition for slashed characters 
% 
\def\slashchar#1{\setbox0=\hbox{$#1$}\dimen0=\wd0% 
\setbox1=\hbox{/}\dimen1=\wd1% 
\ifdim\dimen0>\dimen1%                         
\rlap{\hbox to \dimen0{\hfil/\hfil}}#1\else                                         
\rlap{\hbox to \dimen1{\hfil$#1$\hfil}}/\fi} 
% 
% Redefine \textindent for use in \item 
% 
\def\textindent#1{\noindent\hbox to \parindent{#1\hss}\ignorespaces} 
% 
% Symbols and curves for use in figure captions 
% 
\def\opencirc{\raise1pt\hbox{$\scriptstyle{\bigcirc}$}} 
 
\ifams 
\def\opensqr{\hbox{$\square$}} 
 
\def\opentridown{\hbox{$\triangledown$}}

\else 
\def\opensqr{\vbox{\hrule height.4pt\hbox{\vrule width.4pt height3.5pt 
    \kern3.5pt\vrule width.4pt}\hrule height.4pt}} 
 
\def\opentridown{\raise1pt\hbox{$\scriptstyle\bigtriangledown$}}

           %  These produce the 
                   %  equivalent open character 
           %  to be filled in. 
\fi

% 
% Redefinition of \cases 
% 
\def\m@th{\mathsurround=0pt} 
% 
% Displaystyle now used for first term 
% 
\def\cases#1{% 
\left\{\,\vcenter{\normalbaselines\openup1\jot\m@th% 
     \ialign{$\displaystyle##\hfil$&\rm\tqs##\hfil\crcr#1\crcr}}\right.}% 
% 
% Original version of cases now called \oldcases 
% 
\def\oldcases#1{\left\{\,\vcenter{\normalbaselines\m@th 
    \ialign{$##\hfil$&\rm\quad##\hfil\crcr#1\crcr}}\right.} 
% 
% Cases with number at end each line (using automatic numbering) 
% 
\def\numcases#1{\left\{\,\vcenter{\baselineskip=15pt\m@th% 
     \ialign{$\displaystyle##\hfil$&\rm\tqs##\hfil 
     \crcr#1\crcr}}\right.\hfill 
     \vcenter{\baselineskip=15pt\m@th% 
     \ialign{\rlap{$\phantom{\displaystyle##\hfil}$}\tabskip=0pt&\en 
     \rlap{\phantom{##\hfil}}\crcr#1\crcr}}} 
\def\ptnumcases#1{\left\{\,\vcenter{\baselineskip=15pt\m@th% 
     \ialign{$\displaystyle##\hfil$&\rm\tqs##\hfil 
     \crcr#1\crcr}}\right.\hfill 
     \vcenter{\baselineskip=15pt\m@th% 
     \ialign{\rlap{$\phantom{\displaystyle##\hfil}$}\tabskip=0pt&\enpt 
     \rlap{\phantom{##\hfil}}\crcr#1\crcr}}\global\eqlett=97 
     \global\advance\countno by 1} 
% 
% for equation numbers instead of \eqno 
% 
\def\eq(#1){\ifaligned\@mp(#1)\else\hfill\llap{{\rm (#1)}}\fi} 
\def\ceq(#1){\ns\ns\ifaligned\@mp\fi\eq(#1)\cr\ns\ns} 
\def\eqpt(#1#2){\ifaligned\@mp(#1{\it #2\/}) 
                    \else\hfill\llap{{\rm (#1{\it #2\/})}}\fi} 
 
% 
% Automatic numbering of equations 
% 
\countno=1 
\def\eqnobysec{\numbysectrue} 
\def\aleq{&\rm(\ifappendix\applett 
               \ifnumbysec\ifnum\secno>0 \the\secno\fi.\fi 
               \else\ifnumbysec\the\secno.\fi\fi\the\countno} 
\def\noaleq{\hfill\llap\bgroup\rm(\ifappendix\applett 
               \ifnumbysec\ifnum\secno>0 \the\secno\fi.\fi 
               \else\ifnumbysec\the\secno.\fi\fi\the\countno} 
\def\@mp{&} 
\def\en{\ifaligned\aleq)\else\noaleq)\egroup\fi\gac} 
\def\cen{\ns\ns\ifaligned\@mp\fi\en\cr\ns\ns} 
\def\enpt{\ifaligned\aleq{\it\char\the\eqlett})\else 
    \noaleq{\it\char\the\eqlett})\egroup\fi 
    \global\advance\eqlett by 1} 
\def\endpt{\ifaligned\aleq{\it\char\the\eqlett})\else 
    \noaleq{\it\char\the\eqlett})\egroup\fi 
    \global\eqlett=97\gac} 
% 
% abbreviations for Institute of Physics Publishing journals 
% 

\def\JPA{{\it J. Phys. A: Math. Gen.}} 
        %1968-87 
   %1988 and onwards 
     %1968--1988 
        %1989 and onwards 

           %1975--1988 
     %1989 and onwards 
 
                 %1990 and onwards 

% 
% Other commonly quoted journals 
% 

\def\AP{{\it Ann. Phys., Lpz.}} 
\def\APNY{{\it Ann. Phys., NY\/}}

\def\NP{{\it Nucl. Phys.}} 
 
\def\PR{{\it Phys. Rev.}} 
\def\PRL{{\it Phys. Rev. Lett.}}

\def\RMP{{\it Rev. Mod. Phys.}}

\def\ZP{{\it Z. Phys.}} 
\headline={\ifodd\pageno{\ifnum\pageno=\firstpage\hfill 
   \else\rrhead\fi}\else\lrhead\fi} 
\def\rrhead{\textfonts\hskip\secindent\it 
    \shorttitle\hfill\rm\folio} 
\def\lrhead{\textfonts\hbox to\secindent{\rm\folio\hss}% 
    \it\aunames\hss} 
\footline={\ifnum\pageno=\firstpage \hfill\textfonts\rm\folio\fi} 
\def\@rticle#1#2{\vglue.5pc 
    {\parindent=\secindent \bf #1\par} 
     \vskip2.5pc 
    {\exhyphenpenalty=10000\hyphenpenalty=10000 
     \baselineskip=18pt\raggedright\noindent 
     \headfonts\bf#2\par}\futurelet\next\sh@rttitle}% 
\def\title#1{\gdef\shorttitle{#1} 
    \vglue4pc{\exhyphenpenalty=10000\hyphenpenalty=10000  
    \baselineskip=18pt  
    \raggedright\parindent=0pt 
    \headfonts\bf#1\par}\futurelet\next\sh@rttitle}  

\def\article#1#2{\gdef\shorttitle{#2}\@rticle{#1}{#2}}  
\def\review#1{\gdef\shorttitle{#1}% 
    \@rticle{REVIEW \ifpbm\else ARTICLE\fi}{#1}} 
\def\topical#1{\gdef\shorttitle{#1}% 
    \@rticle{TOPICAL REVIEW}{#1}} 
\def\comment#1{\gdef\shorttitle{#1}% 
    \@rticle{COMMENT}{#1}} 
\def\note#1{\gdef\shorttitle{#1}% 
    \@rticle{NOTE}{#1}} 
\def\prelim#1{\gdef\shorttitle{#1}% 
    \@rticle{PRELIMINARY COMMUNICATION}{#1}} 
\def\letter#1{\gdef\shorttitle{Letter to the Editor}% 
     \gdef\aunames{Letter to the Editor} 
     \global\lettertrue\ifnum\jnl=7\global\letterfalse\fi 
     \@rticle{LETTER TO THE EDITOR}{#1}} 
\def\sh@rttitle{\ifx\next[\let\next=\sh@rt 
                \else\let\next=\f@ll\fi\next} 
\def\sh@rt[#1]{\gdef\shorttitle{#1}} 
\def\f@ll{} 
\def\author#1{\ifletter\else\gdef\aunames{#1}\fi\vskip1.5pc 
    {\parindent=\secindent   
     \hang\textfonts   
     \ifppt\bf\else\rm\fi#1\par}   
     \ifppt\bigskip\else\smallskip\fi 
     \futurelet\next\@unames} 
\def\@unames{\ifx\next[\let\next=\short@uthor 
                 \else\let\next=\@uthor\fi\next} 
\def\short@uthor[#1]{\gdef\aunames{#1}} 
\def\@uthor{} 
\def\address#1{{\parindent=\secindent 
    \exhyphenpenalty=10000\hyphenpenalty=10000 
\ifppt\textfonts\else\smallfonts\fi\hang\raggedright\rm#1\par}% 
    \ifppt\bigskip\fi} 
\def\jl#1{\global\jnl=#1} 
\jl{0}% 
\def\journal{\ifnum\jnl=1 J. Phys.\ A: Math.\ Gen.\  
        \else\ifnum\jnl=2 J. Phys.\ B: At.\ Mol.\ Opt.\ Phys.\  
        \else\ifnum\jnl=3 J. Phys.:\ Condens. Matter\  
        \else\ifnum\jnl=4 J. Phys.\ G: Nucl.\ Part.\ Phys.\  
        \else\ifnum\jnl=5 Inverse Problems\  
        \else\ifnum\jnl=6 Class. Quantum Grav.\  
        \else\ifnum\jnl=7 Network\  
        \else\ifnum\jnl=8 Nonlinearity\ 
        \else\ifnum\jnl=9 Quantum Opt.\ 
        \else\ifnum\jnl=10 Waves in Random Media\ 
        \else\ifnum\jnl=11 Pure Appl. Opt.\  
        \else\ifnum\jnl=12 Phys. Med. Biol.\ 
        \else\ifnum\jnl=13 Modelling Simulation Mater.\ Sci.\ Eng.\  
        \else\ifnum\jnl=14 Plasma Phys. Control. Fusion\  
        \else\ifnum\jnl=15 Physiol. Meas.\  
        \else\ifnum\jnl=16 Sov.\ Lightwave Commun.\ 
        \else\ifnum\jnl=17 J. Phys.\ D: Appl.\ Phys.\ 
        \else\ifnum\jnl=18 Supercond.\ Sci.\ Technol.\ 
        \else\ifnum\jnl=19 Semicond.\ Sci.\ Technol.\ 
        \else\ifnum\jnl=20 Nanotechnology\ 
        \else\ifnum\jnl=21 Meas.\ Sci.\ Technol.\  
        \else\ifnum\jnl=22 Plasma Sources Sci.\ Technol.\  
        \else\ifnum\jnl=23 Smart Mater.\ Struct.\  
        \else\ifnum\jnl=24 J.\ Micromech.\ Microeng.\ 
   \else Institute of Physics Publishing\  
   \fi\fi\fi\fi\fi\fi\fi\fi\fi\fi\fi\fi\fi\fi\fi 
   \fi\fi\fi\fi\fi\fi\fi\fi\fi} 
\let\abs=\beginabstract 

\let\endabs=\endabstract 
\def\today{\number\day\ \ifcase\month\or 
     January\or February\or March\or April\or May\or June\or 
     July\or August\or September\or October\or November\or 
     December\fi\space \number\year} 
\def\date{\ifppt\noindent\textfonts\rm  
     Date: \today\par\goodbreak\bigskip\fi} 
% 
% Physics Abstracts classification numbers 
% 
 
% 
 
% 
%%%%%%%%%%%%%%%%%%%%%%%%%%%%%%%%%%%%%%%%%%%%%%%%%%%%%%%%%%%% 
%                                                          % 
%  Sections, subsections, etc                              % 
%                                                          % 
%%%%%%%%%%%%%%%%%%%%%%%%%%%%%%%%%%%%%%%%%%%%%%%%%%%%%%%%%%%% 
% 
\def\section#1{\ifppt\ifnum\secno=0\eject\fi\fi 
    \subno=0\subsubno=0\global\advance\secno by 1 
    \gdef\labeltype{\seclabel}\ifnumbysec\countno=1\fi 
    \goodbreak\beforesecspace\nobreak 
    \noindent{\bf \the\secno. #1}\par\futurelet\next\sp@ce} 
\def\subsection#1{\subsubno=0\global\advance\subno by 1 
     \gdef\labeltype{\seclabel}% 
     \ifssf\else\goodbreak\beforesubspace\fi 
     \global\ssffalse\nobreak 
     \noindent{\it \the\secno.\the\subno. #1\par}% 
     \futurelet\next\subsp@ce} 
\def\subsubsection#1{\global\advance\subsubno by 1 
     \gdef\labeltype{\seclabel}% 
     \ifssf\else\goodbreak\beforesubsubspace\fi 
     \global\ssffalse\nobreak 
     \noindent{\it \the\secno.\the\subno.\the\subsubno. #1}\null.  
     \ignorespaces} 
% 
 
% 
%%%%%%%%%%%%%%%%%%%%%%%%%%%%%%%%%%%%%%%%%%%%%%%%%%%%%%%%%%%% 
%                                                          % 
%  Appendices                                              % 
%                                                          % 
%%%%%%%%%%%%%%%%%%%%%%%%%%%%%%%%%%%%%%%%%%%%%%%%%%%%%%%%%%%% 
% 
\def\numappendix#1{\ifappendix\ifnumbysec\countno=1\fi\else 
    \countno=1\figno=0\tabno=0\fi 
    \subno=0\global\advance\appno by 1 
    \secno=\appno\gdef\applett{A}\gdef\labeltype{\seclabel}% 
    \global\appendixtrue\global\numapptrue 
    \goodbreak\beforesecspace\nobreak 
    \noindent{\bf Appendix \the\appno. #1\par}% 
    \futurelet\next\sp@ce} 
\def\numsubappendix#1{\global\advance\subno by 1\subsubno=0 
    \gdef\labeltype{\seclabel}% 
    \ifssf\else\goodbreak\beforesubspace\fi 
    \global\ssffalse\nobreak 
    \noindent{\it A\the\appno.\the\subno. #1\par}% 
    \futurelet\next\subsp@ce} 
\def\@ppendix#1#2#3{\countno=1\subno=0\subsubno=0\secno=0\figno=0\tabno=0 
    \gdef\applett{#1}\gdef\labeltype{\seclabel}\global\appendixtrue 
    \goodbreak\beforesecspace\nobreak 
    \noindent{\bf Appendix#2#3\par}\futurelet\next\sp@ce} 
\def\Appendix#1{\@ppendix{A}{. }{#1}} 
\def\appendix#1#2{\@ppendix{#1}{ #1. }{#2}} 
\def\App#1{\@ppendix{A}{ }{#1}} 
\def\app{\@ppendix{A}{}{}} 
\def\subappendix#1#2{\global\advance\subno by 1\subsubno=0 
    \gdef\labeltype{\seclabel}% 
    \ifssf\else\goodbreak\beforesubspace\fi 
    \global\ssffalse\nobreak 
    \noindent{\it #1\the\subno. #2\par}% 
    \nobreak\subspace\noindent\ignorespaces} 
% 
%%%%%%%%%%%%%%%%%%%%%%%%%%%%%%%%%%%%%%%%%%%%%%%%%%%%%%%%%%%% 
%                                                          % 
%  Acknowledgments, notes added and foreign abstracts      % 
%                                                          % 
%%%%%%%%%%%%%%%%%%%%%%%%%%%%%%%%%%%%%%%%%%%%%%%%%%%%%%%%%%%% 
% 
\def\@ck#1{\ifletter\bigskip\noindent\ignorespaces\else 
    \goodbreak\beforesecspace\nobreak 
    \noindent{\bf Acknowledgment#1\par}% 
    \nobreak\secspace\noindent\ignorespaces\fi} 
\def\ack{\@ck{s}} 
\def\ackn{\@ck{}} 
\def\n@ip#1{\goodbreak\beforesecspace\nobreak 
    \noindent\smallfonts{\it #1}. \rm\ignorespaces} 
\def\naip{\n@ip{Note added in proof}} 
\def\na{\n@ip{Note added}} 
 
% 
%  \resume and \zus in Physics in Medicine and Biology only 
% 
 
% 
 
% 
%%%%%%%%%%%%%%%%%%%%%%%%%%%%%%%%%%%%%%%%%%%%%%%%%%%%%%%%%%%% 
%                                                          % 
%  Tables                                                  % 
%                                                          % 
%%%%%%%%%%%%%%%%%%%%%%%%%%%%%%%%%%%%%%%%%%%%%%%%%%%%%%%%%%% 
% 
 
% 
 
% 
 
\def\tablecont{\topinsert\global\advance\tabno by -1 
    \tablecaption{(continued)}} 
\def\tablecaption#1{\gdef\labeltype{\tablabel}\global\widefalse 
    \leftskip=\secindent\parindent=0pt 
    \global\advance\tabno by 1 
    \smallfonts{\bf Table \ifappendix\applett\fi\the\tabno.} \rm #1\par 
    \smallskip\futurelet\next\t@b} 
\def\t@b{\ifx\next*\let\next=\widet@b 
             \else\ifx\next[\let\next=\fullwidet@b 
                      \else\let\next=\narrowt@b\fi\fi 
             \next} 
\def\widet@b#1{\global\widetrue\global\notfulltrue 
    \t@bwidth=\hsize\advance\t@bwidth by -\secindent}  
\def\fullwidet@b[#1]{\global\widetrue\global\notfullfalse 
    \leftskip=0pt\t@bwidth=\hsize}                   
\def\narrowt@b{\global\notfulltrue} 
\def\align{\catcode`?=13\ifnotfull\moveright\secindent\fi 
    \vbox\bgroup\halign\ifwide to \t@bwidth\fi 
    \bgroup\strut\tabskip=1.2pc plus1pc minus.5pc} 
\def\endalign{\egroup\egroup\catcode`?=12} 
 
% 
% Use \L{#}, \R{#} and \C{#} to specify left, right or centred 
% columns immediately after \table. For example 
% \align\L{#}&&\L{#}\cr gives the preamble for a table with 
% all columns aligned left, \align\L{#}&\C{#}&\R{#}\cr 
% gives a table with 3 columns, the first aligned left, the second 
% centred and the third aligned right. 
% 

% 
%  Rules for tables \br at top and bottom 
%  \mr to separate headings from entries 
% 

% 
 
% 
% Definitions for centring headings over several columns 
% \centre{4}{Results for helium} will centre 
% Results for helium over four columns 
% \crule{4} will produce a rule centred over four columns 
% to go below a centred heading 
% 

% 
 
\catcode`?=13 
\def\lineup{\setbox0=\hbox{\smallfonts\rm 0}% 
    \digitwidth=\wd0% 
    \def?{\kern\digitwidth}% 
    \def\\{\hbox{$\phantom{-}$}}% 
    \def\-{\llap{$-$}}} 
\catcode`?=12 
% 
% Macros for two parts of a table of equal width side by side 
% \table{caption}[w] 
% \sidetable{first part}{second part} 
% \endtable 
% Use \table preamble for tables of 31picas width 
% 
\def\sidetable#1#2{\hbox{\ifppt\hsize=18pc\t@bwidth=18pc 
                          \else\hsize=15pc\t@bwidth=15pc\fi 
    \parindent=0pt\vtop{\null #1\par}% 
    \ifppt\hskip1.2pc\else\hskip1pc\fi 
    \vtop{\null #2\par}}}  
\def\lstable#1#2{\everypar{}\tempval=\hsize\hsize=\vsize 
    \vsize=\tempval\hoffset=-3pc 
    \global\tabno=#1\gdef\labeltype{\tablabel}% 
    \noindent\smallfonts{\bf Table \ifappendix\applett\fi 
    \the\tabno.} \rm #2\par 
    \smallskip\futurelet\next\t@b} 
\def\inctabno{\global\advance\tabno by 1} 
% 
%%%%%%%%%%%%%%%%%%%%%%%%%%%%%%%%%%%%%%%%%%%%%%%%%%%%%%%%%%%% 
%                                                          % 
%  Figures                                                 % 
%                                                          % 
%%%%%%%%%%%%%%%%%%%%%%%%%%%%%%%%%%%%%%%%%%%%%%%%%%%%%%%%%%%% 
% 
 
% 
 
% 
\def\figure#1{\figc@ption{#1}\bigskip} 
\def\figc@ption#1{\global\advance\figno by 1\gdef\labeltype{\figlabel}% 
   {\parindent=\secindent\smallfonts\hang 
    {\bf Figure \ifappendix\applett\fi\the\figno.} \rm #1\par}} 
% 
%%%%%%%%%%%%%%%%%%%%%%%%%%%%%%%%%%%%%%%%%%%%%%%%%%%%%%%%%%%% 
%                                                          % 
%  Reference lists                                         % 
%                                                          % 
%%%%%%%%%%%%%%%%%%%%%%%%%%%%%%%%%%%%%%%%%%%%%%%%%%%%%%%%%%%% 
% 
\def\refHEAD{\goodbreak\beforesecspace 
     \noindent\textfonts{\bf References}\par 
     \let\ref=\rf 
     \nobreak\smallfonts\rm} 
\def\numreferences{\refHEAD\parindent=30pt 
     \everypar{\hang\noindent\frenchspacing\rm} 
     \secspace} 
\def\rf#1{\par\noindent\hbox to 21pt{\hss #1\quad}\ignorespaces} 
% 
 
% 
 
% 
% reference to a journal article in numerical system 
% 
 
% 
% reference to a book or report in numerical system 
% 
 
% 
%%%%%%%%%%%%%%%%%%%%%%%%%%%%%%%%%%%%%%%%%%%%%%%%%%%%%%%%%%%% 
%                                                          % 
%  Theorems, lemmas, etc                                   % 
%                                                          % 
%%%%%%%%%%%%%%%%%%%%%%%%%%%%%%%%%%%%%%%%%%%%%%%%%%%%%%%%%%%% 
% 

% 
% NB \note#1 is used to give a Note (as opposed to a paper or letter) 
% in PMB therefore use commands \notes#1 for numbered Note 
% instead of \note  
% 

% 
\catcode`\@=12 
% 
% Parameter values for `Preprint' style  
% 
 
% 
% Parameter values for `Journal' style  
% 
\def\jnlstyle{\pptfalse\headsize{14}{18}% 
\textsize{10}{12}% 
\smallsize{8}{10} 
\textind=16pt} 
% 
% Parameter values for `Eleven point' style  
% 
 
% 
% Parameter values for `Large size' style  
% 
 
% 
\parindent=\textind 
%%%%%%%%%%%%%%%%%%%%%%%%%%%%%%%%%%%%%%%%%%%%%%%%%%%%%%%%%%%%%%%%%%%%%%%%%%%%%%%
\input xref
\input epsf
\eqnobysec
%
%\def\ps{\noalign{\vskip3pt}}
% positive space in tables, matrices 
% \ns for negative space
\def\received#1{\insertspace 
     \parindent=\secindent\ifppt\textfonts\else\smallfonts\fi 
     \hang{Received #1}\rm } 
%\def\appendix{\goodbreak\beforesecspace 
%     \noindent\textfonts{\bf Appendix}\secspace} 
%\headline={\ifodd\pageno{\ifnum\pageno=\firstpage\titlehead
%   \else\rrhead\fi}\else\lrhead\fi} 
%\def\lpm#1#2{LPM-#1-LT#2}
%\def\figure#1{\global\advance\figno by 1\gdef\labeltype{\figlabel}% 
%   {\parindent=\secindent\smallfonts\hang 
%    {\bf Figure \ifappendix\applett\fi\the\figno.} \rm #1\par}} 
%\def\endtable{\parindent=\textind\textfonts\rm\bigskip} 
%\def\rrhead{\textfonts\hskip\secindent\it 
%    \shorttitle\hfill\folio} 
%\def\lrhead{\textfonts\hbox to\secindent{\folio\hss}% 
%   \it\aunames\hss} 
%\footline={\ifnum\pageno=\firstpage
%\smallfonts cond-mat/\hfil\textfonts\folio\fi}   
%\def\titlehead{\smallfonts 
%\hfil\lpm{}{}} 
\jnlstyle
%\pptstyle
\firstpage=1
\pageno=1

\jnlstyle

\jl{1}

\title{Conformal profiles in the Hilhorst--van Leeuwen
model}[Conformal profiles in the Hilhorst--van Leeuwen model]

\author{D Karevski\dag, L Turban\dag\ and F Igl\'oi\dag\ddag\S}[D Karevski, L
Turban and F Igl\'oi]

\address{\dag Laboratoire de Physique des
Mat\'eriaux, Universit\'e Henri Poincar\'e (Nancy~I),
BP~239,  F--54506~Vand\oe uvre l\`es Nancy Cedex, France}

\address{\ddag Research Institute for Solid State Physics and
Optics, PO Box 49, H--1525 Budapest, Hungary}

\address{\S Institute of Theoretical Physics, Szeged University, H--6720
Szeged, Hungary}

\received{22 December 1999}

\abs
We study the critical energy and magnetization profiles for the Ising quantum
chain with a marginal extended surface perturbation of the form $A/y$, $y$
being the distance from the surface (Hilhorst-van Leeuwen model). For weak
local couplings, $A<A_{\rm c}$, the model displays a continuous surface
transition with $A$-dependent exponents, whereas, for $A>A_{\rm c}$, there is
surface order at the bulk critical point. If conformal invariance is assumed
to hold with such marginal perturbations, it predicts 
conformal profiles with the same scaling form as for the unperturbed quantum
chain, with marginal surface exponents replacing the
unperturbed ones. The results of direct analytical and numerical calculations
of the profiles confirm the validity of the conformal expressions in the
regimes of second- and first-order surface transitions.  
\endabs 
\date 
\vglue 1.5cm

\section{Introduction}

The methods of conformal invariance have been extensively used in
two-dimensional (2D) critical systems, first to study the bulk critical
properties~\cite{belavin84} but also, very soon after, to explore the local
critical behaviour at surfaces~\cite{cardy84} or near line
defects~\cite{turban85}. Although translation invariance is broken in the
transverse direction for such defective systems, conformal invariance is
preserved. In the case of the line defect, the spectrum-generating algebra
was identified as the $U(1)$ Kac-Moody algebra~\cite{henkel87}. More
surprisingly, some aspects of conformal invariance, such as gap-exponent
relations, seem to hold for marginal extended perturbations, provided the
shape of the perturbation is properly transformed~[5--10].

Extended perturbations were first considered at the surface of the
semi-infinite  2D Ising model by Hilhorst and van Leeuwen
(HvL)~\cite{hilhorst81}. This problem, as well as its extensions to linear and
radial defects in the bulk, became an active field of research during the
years that followed~[12--20]. Different approaches, either numerical or
analytical, have been used to elucidate the influence of such perturbations
on the local critical behaviour (see~\cite{igloi93} for a review).

Surface extended perturbations correspond to smoothly varying
inhomogeneties for which the nearest-neighbour coupling, at a
distance $y$ from the surface, deviates from its bulk value by a quantity
$Ay^{-\kappa}$.
When the decay exponent $\kappa$ is equal to the scaling dimension
of the couplings, the perturbation is marginal and the local critical
exponents vary continuously with the perturbation amplitude
$A$~[22--24]. For a sufficiently strong
enhancement of the perturbation amplitude $A$, the surface remains ordered
at the bulk critical point. When $\kappa$ is larger than the marginal
value, the perturbation  decreases sufficiently fast and the critical
properties remain unchanged. On the other hand, for a value of
$\kappa$ smaller than the critical one, the system is driven to a new surface
fixed point, at which power laws are replaced by stretched exponentials for
$A<0$ whereas the surface remains ordered at criticality when $A>0$.

In the present paper, we consider the off-diagonal profiles~\cite{turban97} of
the energy and magnetization densities for the 2D Ising model
with a marginal HvL surface extended defect. We work in a
strip geometry resulting from a logarithmic conformal mapping of the
half-space. We  treat the problem in the extreme anisotropic
limit~\cite{kogut79}, where the 2D classical Ising model corresponds to a
one-dimensional (1D) Ising quantum chain in a transverse field. The energy
density profiles are obtained analytically, working in the continuum limit, 
whereas the magnetization profiles are studied numerically. The  results are
compared with the conformal predictions in the different regimes.

The paper is organised as follows. In section~2, we review briefly the
conformal results for the surface correlation functions in the presence of a
marginal HvL perturbation and  the resulting off-diagonal profiles in the
strip geometry. In section~3, we present the HvL model in
the extreme anisotropic limit and the diagonalization of the associated
Hamiltonian in the continuum limit. The off-diagonal profiles are calculated
in section 4. The results are discussed in section 5.

\section{Marginal extended perturbations and conformal invariance}

Following~\cite{turban93} let us first consider the behaviour,
under the infinitesimal special conformal transformation 
$z^\prime=z+\epsilon z^2$, of a marginal inhomogeneous coupling of the form
$\Delta(\rho,\theta)=A\rho^{-y_\chi}f(\theta)$, conjugate to the field $\chi$. We
use polar coordinates, $z=\rho\exp(\i\theta)$, and $y_\chi$ is the scaling
dimension of $\Delta$. Up to $\Or(\epsilon)$ we have
$$
\rho^\prime=\rho+\epsilon\rho^2\cos\theta
\qquad\theta^\prime=\theta+\epsilon\rho\sin\theta 
\label{e2.01}
$$
so that the half-plane, with $0\leq\theta\leq\pi$, is mapped onto itself.
The local dilatation factor is $b(\rho,\theta)=\vert\d z^\prime/\d
z\vert^{-1}=1-2\epsilon\rho\cos\theta$ and leads to the following transformed
perturbation:  
$$
\fl
\Delta(\rho^\prime,\theta^\prime)=(1-2\epsilon\rho\cos\theta)^{y_\chi}A{f(\theta)\over
\rho^{y_\chi}}=\left[1-\epsilon\rho\sin\theta\left(y_\chi\cot\theta+{\d\ln
f\over\d\theta}\right)\right]A{f(\theta^\prime)\over{\rho^\prime}^{y_\chi}}\,.
\label{e2.02}
$$
The perturbation is invariant when the coefficient of $\epsilon$ vanishes,
i.e., when $\Delta(\rho,\theta)=A\vert\rho\sin\theta\vert^{-y_\chi}$, which is
just the form of the marginal HvL perturbation. 

If conformal invariance holds with such a marginal perturbation, the
two-point correlation functions satisfy the same set of partial differential
equations as the unperturbed ones~\cite{cardy84}. 

The transformation properties of the two-point functions under the
infinitesimal special conformal transformation imply the following scaling
behaviour at criticality~\cite{cardy84}:  
$$
\fl
G_{\phi\phi}(x_{1}-x_{2},y_{1},y_{2})=(y_{1}y_{2})^{-x_\phi}g_\phi(\omega)
\qquad \omega=\frac{(x_{1}-x_{2})^{2}+(y_{1}-y_{2})^{2}}{y_{1}y_{2}}
\label{e2.1}
$$
where $x$ ($y$) is the direction parallel (perpendicular) to the surface and
$x_\phi$ the bulk scaling dimension of the field $\phi$. Using polar
coordinates and taking the limit $\rho_{1}\gg\rho_{2}$, simple scaling
considerations lead to 
$$
G_{\phi\phi}(\rho_{1},\rho_{2},\theta_{1},\theta_{2})\sim(\rho_{1}\rho_{2})^{-x_
\phi}
\left(\frac{\rho_{1}}{\rho_{2}}\right)^{-x_\phi^{\rm s}}
(\sin\theta_{1}\sin\theta_{2})^{x_\phi^{\rm s}-x_\phi}
\label{e2.2}
$$
where $x_\phi^{\rm s}$ is the surface scaling dimension of $\phi$. Using the
logarithmic conformal transformation, $w=(L/\pi)\ln z$, with $z=x+\i
y=\rho\exp(\i\theta)$ and $w=u+\i v$, the half-plane, $y>0$, with uniform
boundary conditions is mapped onto a strip ($-\infty<u<+\infty$, $0<v<L$) with
the same boundary conditions on both edges. The correlation functions in the
two geometries are related under the conformal mapping so that
$$
G_{\phi\phi}(w_{1},w_{2})
=[b(z_{1})b(z_{2})]^{x_\phi}G_{\phi\phi}(z_{1},z_{2})
\label{e2.3}
$$
where $b(z)=|{\rm d}w/{\rm d}z|^{-1}$ is the local dilatation factor. Thus
one obtains the well known result for the two-point function
in the strip geometry~\cite{cardy83}
$$
\fl
G_{\phi\phi}(u_{1}-u_{2},v_{1},v_{2})\sim\!\!
\left(\frac{\pi}{L}\right)^{2x_\phi}\left[\sin\left(\frac{\pi
v_{1}}{L}\right)\sin\left(\frac{\pi v_{2}}{L}\right)\right]^{x_\phi^{\rm
s}-x_\phi}\!\!\!\!\exp\left[-\frac{\pi}{L}x_\phi^{\rm
s}(u_{1}-u_{2})\right] \label{e2.4}
$$
when $u_{1}\gg u_{2}$. 

This result is expected to apply with a marginal
extended perturbation, provided the shape of the perturbation is also
transformed according to the logarithmic mapping. Details will be given in
the next section. Then $x_\phi^{\rm s}$ is the continuously varying surface
scaling dimension of $\phi$ associated with the marginal defect.

In the strip geometry, the row-to-row transfer operator ${\cal T}$ of the
isotropic classical system can be used  to expand the correlation function
$G_{\phi\phi}$ over its eigenstates $\vert n\rangle$ such that ${\cal
T}\vert n\rangle=\exp(-E_n)\vert n\rangle$. This leads to
$$
\fl
G_{\phi\phi}(u_{1}-u_{2},v_{1},v_{2})=\sum_{n>0}
\langle 0\vert\hat\phi(v_{1})\vert n\rangle\langle
n\vert\hat\phi(v_{2})\vert 0\rangle \exp[-(E_n-E_0)(u_{1}-u_{2})]
\label{e2.5}
$$
where $\vert 0\rangle$ is the eigenstate corresponding to the largest
eigenvalue of the transfer operator and $\hat\phi$ the operator associated
with the classical field $\phi$. In the limit $u_{1}\gg u_{2}$, the sum is
dominated by the eigenstate $\vert\phi\rangle$ with the largest eigenvalue
$\exp(-E_\phi)$ and a non-vanishing matrix element $\langle
0\vert\hat\phi\vert\phi\rangle$ so that finally one obtains:
$$
\fl G_{\phi\phi}(u_{1}-u_{2},v_{1},v_{2})\simeq
\langle 0\vert\hat\phi(v_{1})\vert\phi\rangle\langle
\phi\vert\hat\phi(v_{2})\vert
0\rangle \exp[-(E_\phi-E_0)(u_{1}-u_{2})]\,.
\label{e2.6}
$$
A comparison to equation~\ref{e2.4} allows us to deduce the off-diagonal
profile for $\phi$ in the strip geometry~\cite{turban97}:
$$
\phi_{od}(v)=\langle 0\vert\hat\phi(v)\vert\phi\rangle\sim
\left(\frac{\pi}{L}\right)^{x_\phi}\left[\sin\left(\frac{\pi
v}{L}\right)\right]^{x_\phi^{\rm s}-x_\phi}\,.
\label{e2.7}
$$
One also recovers the gap-exponent relation~\cite{cardy83}
$$
E_\phi-E_0=\frac{\pi}{L}x_\phi^{\rm s}
\label{e2.8}
$$
which is valid under this form for the isotropic strip.

Off-diagonal profiles contain singular contributions only, which facilitates
the scaling analysis in numerical calculations. Furthermore, an external
symmetry-breaking field is not needed to obtain a non-vanishing order
parameter density. These profiles also give an information about the surface
critical behaviour since equation~\ref{e2.7} with $v\ll L$ leads to
$\phi_{od}(v)\sim L^{-x_\phi^{\rm s}}$.

One may note that the  diagonal profile
$$
\phi(v)\sim\left[\sin\left(\frac{\pi v}{L}\right)\right]^{-x_\phi}
\label{e2.9}
$$
obtained in~\cite{burkhardt85} for the order  parameter, with
symmetric fixed boundary conditions, is formally recovered by
setting $x_\phi^{\rm s}=0$ in equation~\ref{e2.7}, i.e., the scaling
dimension of the order parameter at the extraordinary surface
transition.

\section{Diagonalization of the HvL Ising quantum chain in the continuum
limit}

Let us start with the classical HvL Ising model on a semi-infinite
square lattice with nearest-neighbour couplings
$$
\Delta K_2(y)=K_2(y)-K_2(\infty)=Ay^{-\kappa}\qquad K_2(\infty)=K_2
\label{e3.1}
$$
in the direction perpendicular to the surface and constant
couplings, $K_1$, in the direction parallel to the surface. Hereafter we
consider the marginal situation where $\kappa=1/\nu=1$ for the 2D Ising
model. Under the logarithmic conformal mapping, the marginal perturbation in
the  half-plane transforms into 
$$
\Delta K_2(v)=A\left[\frac{L}{\pi}\sin\left(\frac{\pi
v}{L}\right)\right]^{-1} \label{e3.2}
$$
in the strip geometry where $0<v<L$.

The spectrum of the row-to-row transfer matrix has been obtained
analytically~\cite{burkhardt90} in the extreme anisotropic limit where
$K_2\to0$, $K_1\to\infty$ while $K_2/K_1^*=\lambda$ remains
constant~\cite{kogut79}. Here $K_1^*$ is the dual coupling such that $\tanh
K_1^*=\exp(-2K_1)$ and self-duality allows to locate the critical point at
$\lambda_{\rm c}=1$. The amplitude $A\to0$, too, and thus can be
parametrized as $A=\alpha K_1^*$.

In the extreme anisotropic limit, the fluctuations are anisotropic with a
correlation length ratio given by~\cite{barber84}
$$
{\xi_2\over\xi_1}={\cosh(2K_2)\over\cosh(2K_1)}\simeq2K_1^*
\label{e3.3}
$$
Isotropy is restored by rescaling the lattice parameter $a_1$ in the
time direction to $a_1=2K_1^*$ measured in units of $a_2$ in the transverse
direction. The row-to-row transfer operator, which can be written as ${\cal
T}_{a_1}=\exp(-2K_1^*{\cal H})$, has to be applied $1/a_1$ times in order to
ensure a transfer by one unit of $a_2$ on the rescaled isotropic system.
Thus one may identitfy the transfer operator ${\cal T}$ of the isotropic
system to $\exp(-{\cal H})$ where ${\cal H}$ is the Hamiltonian of the Ising
quantum chain in a transverse field such that ${\cal H}\vert n\rangle=E_n\vert
n\rangle$.

The critical Hamiltonian reads
$$
{\cal H}=-\case12\left[\sum_{l=1}^{L-1}\lambda_{l}\sigma_{l}^{x}
\sigma_{l+1}^{x}+\sum_{l=1}^{L}\sigma_{l}^{z}\right]
\label{e3.4}
$$
where the $\sigma$s are the Pauli spin operators, and the couplings
$\lambda_{l}$ are given by
$$
\lambda_{l}=1+\alpha \left[\frac{L}{\pi}\sin\left({\pi l\over
L}\right)\right]^{-1}\,.
\label{e3.5}
$$

After a Jordan-Wigner transformation~\cite{jordan28}, the
Hamiltonian~\ref{e3.4} becomes a quadratic form in fermion creation and
annihilation operators. It is diagonalized through a canonical
transformation~\cite{lieb61,pfeuty70} leading to
$$
{\cal H}=\sum_{k}\Lambda_{k}\left(\eta_{k}^{\dagger}\eta_{k}
-\case12\right)\,.
\label{e3.6}
$$
The $\eta_{k}^{\dagger}$s ($\eta_{k}$s) are diagonal fermion creation
(anihilation) operators and the $\Lambda_{k}$s are the energies of
the fermionic excitations. They satisfy the following set of
equations:
$$
({\bss{A}+\bss{B}}){\bphi}_k=\Lambda_k{\bpsi}_k\qquad
({\bss{A}-\bss{B}}){\bpsi}_k=\Lambda_k{\bphi}_k
\label{e3.7}
$$
where ${\bphi}_k$ and ${\bpsi}_k$ are normalized eigenvectors. Working with
the ${\bpsi}_k$s such that $({\bss{A}+\bss{B})(\bss{A}-\bss{B}}){\bpsi}_k=
\Lambda_k^{2}{\bpsi}_k$ and changing $\psi_k(l)$ into $(-1)^l \psi_k(l)$, one
is led to the eigenvalue problem,
$$
\fl
-\lambda_{l-1}\psi_k(l-1)+[\lambda_l^2+1]\psi_k(l)-\lambda_l\psi_k(l
+1)=\Lambda_k^{2}\psi_k(l)\qquad l=1,L
\label{e3.8}
$$
with the boundary conditions
$$
\psi_k(0)=0\qquad
\lambda_l\psi_k(L)-\psi_k(L+1)=0\,.
\label{e3.9}
$$
To study the low-energy spectrum, one may solve the finite-difference
equation~\ref{e3.8}, with $\lambda_l$ given by~\ref{e3.5}, by working in the
continuum limit~\cite{burkhardt90}. To second order in $1/L$, using the
continuum variable $z=\pi l/L$, the difference equation is replaced by the
differential equation
$$
\psi_k^{\prime\prime}(z)+\left[\left(\frac{L\Lambda_k}{\pi}\right)^2-\frac{\alpha^{2
}-\alpha\cos z}{\sin^2z}\right]\psi_k(z)=0
\label{e3.10}
$$
with the boundary conditions
$$
\psi_k(z)\big\vert_{z=0}=0 \qquad \psi_k^\prime(z)-\left.\frac{\alpha}{\sin
z}\psi_k(z)\right\vert_{z=\pi}=0\,.
\label{e3.11}
$$
Using the change of function
$\psi_k(z)=\cos^{-\alpha}(z/2)\sin^{1-\alpha}(z/2)f_k(t)$
where $t=\sin^2(z/2)$, \ref{e3.11} becomes a hypergeometric differential
equation. Its general solution is given by \cite{burkhardt90}
$$
f_k(t)=C_1\,F(a,b;c;t)+C_2\,t^{1-c}F(a-c+1,b-c+1;2-c;t)
\label{e3.12}
$$
where
$$
a=-\alpha+\frac{1}{2}+\frac{\Lambda_k L}{\pi}
\qquad b=-\alpha+\frac{1}{2}-\frac{\Lambda_kL}{\pi}
\qquad c=\case32-\alpha\,.
\label{e3.13}
$$

Due to the boundary conditions in equation~\ref{e3.11}, one has to
separately consider two different regimes:

\item{(i)} When $\alpha<\case12$, the condition at $z=0$ imposes $C_2=0$ while
the quantization of the spectrum,
$$
\Lambda_k=\frac{\pi}{L}\left(\frac{1}{2}-\alpha+k\right)\qquad
k=0,1,2,\dots
\label{e3.14}
$$
follows from the condition at $z=\pi$. The spectrum exhibits a conformal
tower-like structure but, to our knowledge, until now no underlying algebra
has been identified. Using the gap-exponent relation in
equation~\ref{e2.8}, one deduces the critical dimension of the
magnetization from $E_{\sigma}-E_0=\Lambda_0$ and the critical
dimension of the energy density  from
$E_{\epsilon}-E_0=\Lambda_0+\Lambda_{1}$. Thus we have
$$
x_{\rm m}^{\rm s}=\case12-\alpha\qquad
x_{\rm e}^{\rm s}=2-2\alpha\,.
\label{e3.15}
$$
These values are in agreement with the result obtained on a semi-infinite
system for the magnetic exponent~\cite{hilhorst81} and through
finite-size scaling on a strip for both exponents~\cite{berche90}, in the
extreme anisotropic limit.\hfill\break
Using equations~(3.12)--(3.14), one obtains the eigenfunctions
$$
\psi_k(z)\sim\frac{1}{L^{1/2}}\frac{\sin^{1-\alpha}(z/2)}{\cos^{\alpha}(z/2)}
\frac{k!}{(3/2-\alpha)_k}P_k^{(\frac{1}{2}-\alpha,-\frac{1}{2}-\alpha)}[\cos
z]\,,
\label{e3.16}
$$
where $P_k^{(\beta,\gamma)}[z]$ is the Jacobi polynomial of degree
$k$~\cite{abramowitz70}. The prefactor, $1/L^{1/2}$, follows from
the normalization.

\item{\hglue-1mm(ii)} When $\alpha>\case12$, the boundary conditions lead to
$C_1=0$ and $$
\fl
\Lambda_0=\Or(L^{-2\alpha})
\qquad\Lambda_k=\frac{\pi}{L}\left(\frac{1}{2}+\alpha+k-1\right)\qquad
k=1,2,3,\dots\,.
\label{e3.17}
$$
Again a tower-like structure appears. One has to notice that the differential
equa\-tion \ref{e3.10} together with~\ref{e3.11} gives $\Lambda_0=0$
since, in the continuum limit, the excitation spectrum is calculated to the
first order in $1/L$. But, as shown in~\cite{burkhardt90}, the correct
$\Or(L^{-2\alpha})$ behaviour can be extracted from the difference
equation~\ref{e3.8} using the trial eigenvector
$\psi_0(l)\sim\tan^{\alpha}(\pi l/2L)$ given by~\ref{e3.10} with
$\Lambda_0=0$.\hfill\break
The anomalous behaviour of the first excitation is linked
to the persistence of a spontaneous surface magnetization at the critical
point for $\alpha>\case12$~\cite{blote83,peschel84}. The surface order is due
to the enhancement of the couplings near the surface which overcompensates
the missing bond. The surface displays a first-order transition at the
bulk critical point with a jump of the surface magnetization to zero when
$\lambda<1$ since, considering the classical picture, the 1D
surface cannot remain ordered when the bulk is disordered.\hfill\break
{F}rom the gap-exponent relation~\ref{e2.8}, the following critical indices are
deduced:
$$
x_{\rm m}^{\rm s}=0\qquad
x_{\rm e}^{\rm s}=\frac{1}{2}+\alpha\,.
\label{e3.18}
$$
The eigenfunctions read
$$
\fl
\eqalign{
&\psi_0(z)\sim\frac{1}{L^{\alpha}}\tan^{\alpha}\left(\frac{z}{2}\right)\cr
&\psi_{k\ge 1}(z)\sim\frac{1}{L^{1/2}}\tan^{\alpha}\!\left(\frac{z}{2}\right)
\cos^{1+2\alpha}\!\left(\frac{z}{2}\right)
F\left[-k+1,k+2\alpha;\frac{1}{2}+\alpha;\sin^2\!
\left(\frac{z}{2}\right)\right]}
\label{e3.19}
$$
where the hypergeometric function, $F$, reduces to a polynomial of order
$k-1$. The anomalous normalization of $\psi_0(z)$ is a consequence of the
spontaneous surface order.

In analogy with surface transitions in homogeneous systems, (i) is a regime
of ordinary surface transition. For (ii) we use the name of first-order
surface transition.

\section{Off-diagonal conformal profiles}
\subsection{Energy-density profiles}
First we consider the energy-density profile for which we derive exact
expressions. The off-diagonal profile is given by the matrix element
$e(l)=\langle0\vert\sigma_l^{z}\vert\epsilon\rangle$  where
$\vert\epsilon\rangle=\eta_{1}^{\dagger}\eta_0^{\dagger}\vert0\rangle$ is the
lowest excited state with an even number of excitations. Writing $\sigma^{z}$
in terms of diagonal fermions, in the continuum limit, one obtains
\cite{berche90}
$$
e(z)=\psi_0(z)\psi_{1}(\pi-z)+\psi_0(\pi-z)\psi_{1}(z)\,.
\label{e4.1}
$$

%%%%% figure 1 %%%%%%%%%%%%%%%%%%%%%%%%%%%%%%%%%%%%%%%%%%%%%%%%
{\par\begingroup\medskip
\epsfxsize=7.5truecm
\topinsert
\centerline{\epsfbox{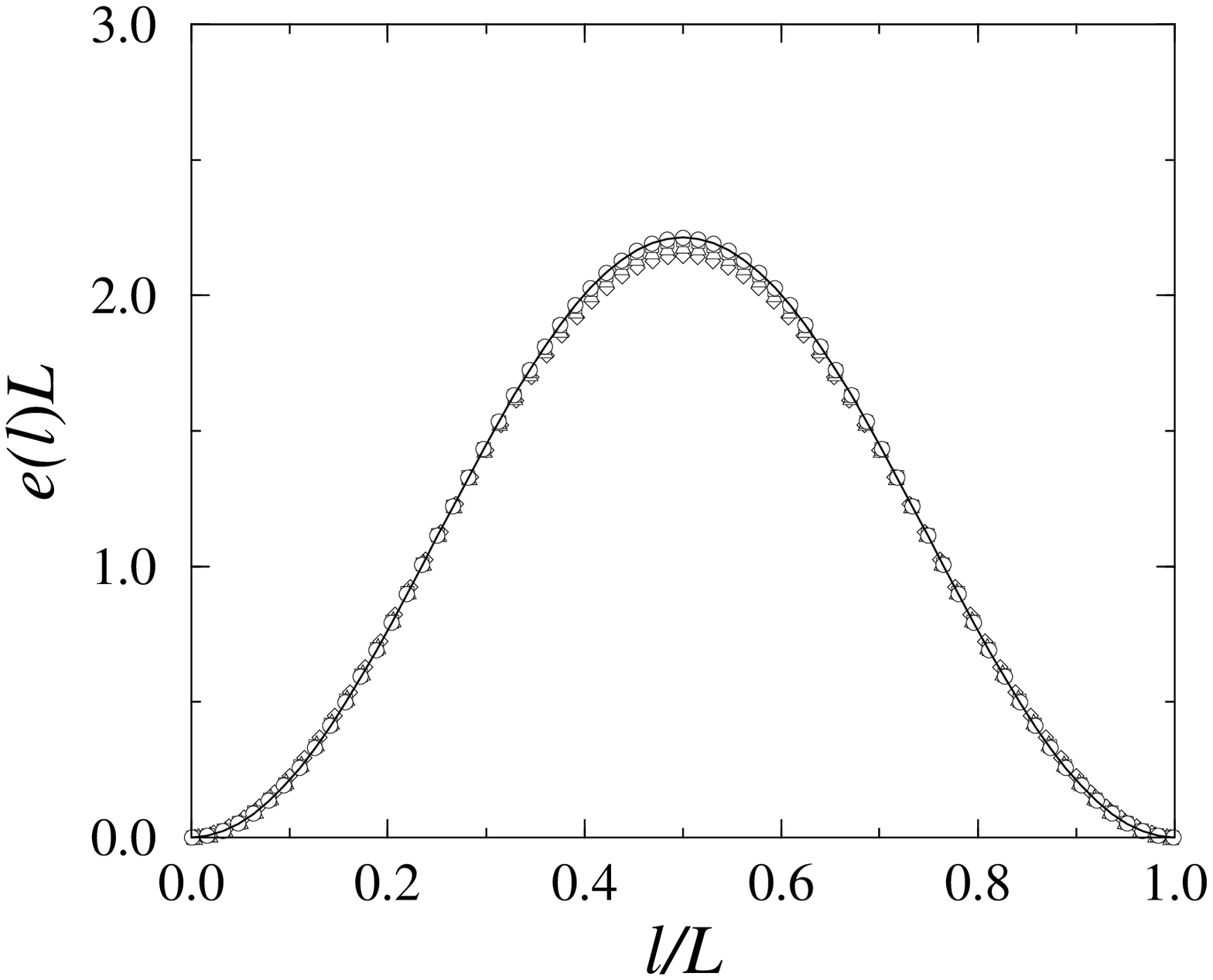}}
\vglue5mm
\figure{Rescaled energy-density profile $e(z)$ in the regime of ordinary surface
transition with $\alpha=-\case12$ for chain sizes $L=2^6$ (diamond), 
$2^7$ (triangle), $2^8$ (square) and
$2^{9}$ (circle). The conformal profile (full curve) is shown for comparison.
\label{fig1}} \endinsert
\endgroup
\par}
%%%%%%%%%%%%%%%%%%%%%%%%%%%%%%%%%%%%%%%%%%%%%%%%%%%%%
%%%%% figure 2 %%%%%%%%%%%%%%%%%%%%%%%%%%%%%%%%%%%%%%%%%%%%%%%%
{\par\begingroup\medskip
\epsfxsize=7.5truecm
\topinsert
\centerline{\epsfbox{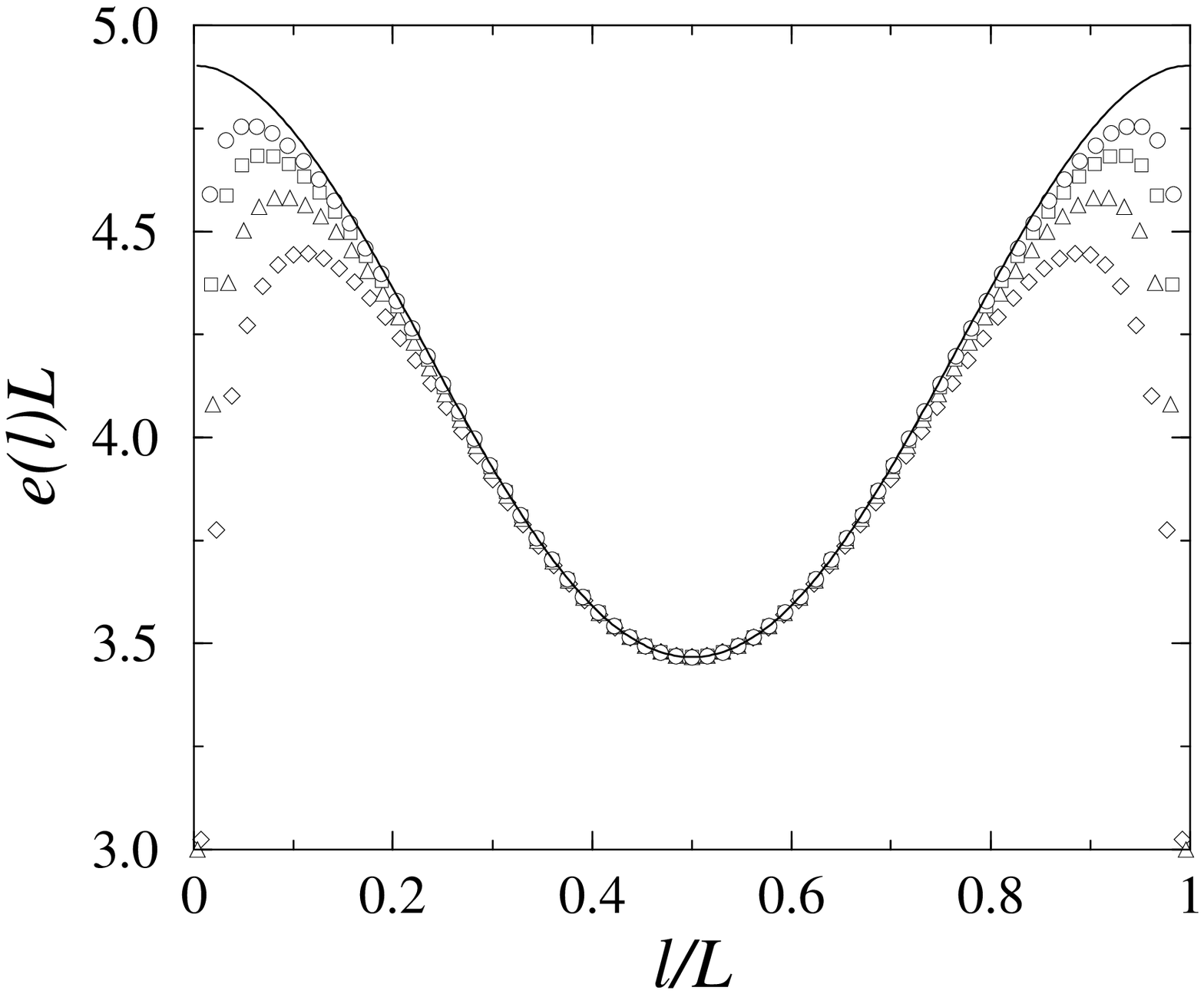}}
\vglue5mm
\figure{Rescaled energy-density profile $e(z)$ in the regime of first-order surface
transition with $\alpha=1$ for chain sizes $L=2^6$ (diamond), 
$L=2^7$ (triangle), $2^8$ (square) and $2^9$ (circle). The
result obtained in the continuum approximation (full curve) is shown for
comparison.   
\label{fig2}}
\endinsert
\endgroup
\par}
%%%%%%%%%%%%%%%%%%%%%%%%%%%%%%%%%%%%%%%%%%%%%%%%%%%%%

%%%%% figure 3 %%%%%%%%%%%%%%%%%%%%%%%%%%%%%%%%%%%%%%%%%%%%%%%%
{\par\begingroup\medskip
\epsfxsize=7.5truecm
\midinsert
\centerline{\epsfbox{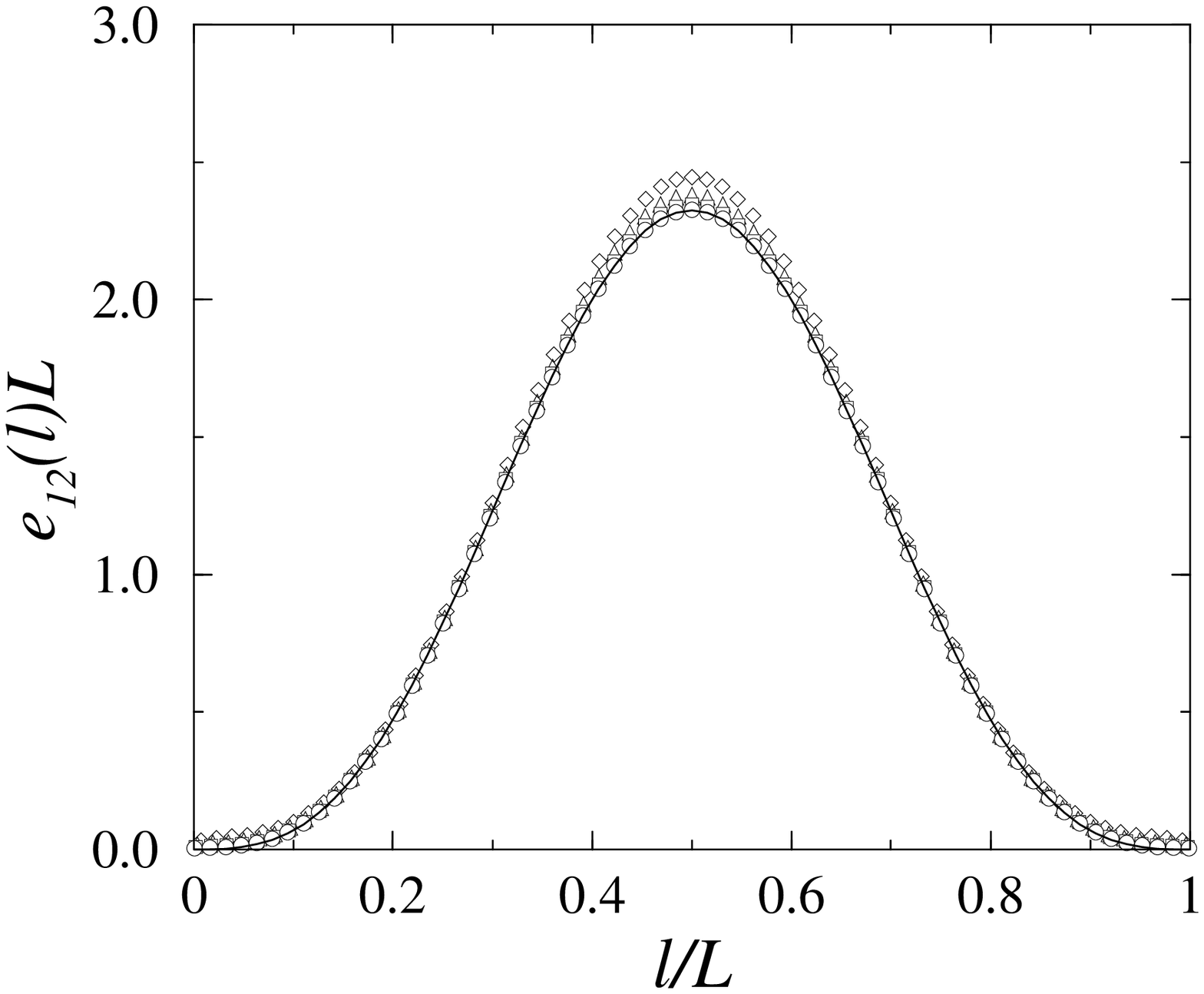}}
\vglue5mm
\figure{Rescaled energy-density profile $e_{12}(z)$ in the regime of first-order surface
transition with $\alpha=1$ for chain sizes $L=2^6$ (diamond), 
$L=2^7$ (triangle), $2^8$ (square) and $2^9$ (circle). The
conformal profile (full curve) is shown for comparison.  
\label{fig3}}
\endinsert
\endgroup
\par}
%%%%%%%%%%%%%%%%%%%%%%%%%%%%%%%%%%%%%%%%%%%%%%%%%%%%%

\item{(i)} At the ordinary surface transition, $\alpha<\case12$, using
equation~\ref{e3.16} with $k=0$ and $k=1$, a straightforward calculation
leads to the following expression:
$$
e(z)=\frac{A}{L}(\sin z)^{1-2\alpha}
\label{e4.2}
$$
where $A$ is a constant amplitude. This expression agrees with the conformal
result in equation~\ref{e2.7} since $z=\pi l/L$, the bulk scaling dimension
of the energy is $x_{\rm e}=1$ for the Ising model and the
surface one is $x_{\rm e}^{\rm s}=2-2\alpha$ according to~\ref{e3.15}.

\item{\hglue-1mm(ii)} At the border line, $\alpha=\case12$, the surface
magnetization vanishes logarithmically~\cite{blote83} and the first excitation
scales normally as $1/L$.
One may use equation~\ref{e4.1} to calculate the energy profile. In this case
both~\ref{e3.16} and~\ref{e3.19} give the same solution. In
particular $\psi_0(z)\sim  L^{-1/2}\tan^{1/2}(z/2)$ and
$\psi_{1}(z)\sim L^{-1/2}\tan^{1/2}(z/2)\cos^2(z/2)$. Inserting these
eigenvectors into~\ref{e4.1} one obtains a constant profile,
$$
e(z)=\frac{D}{L}
\label{e4.3}
$$
in the continuum limit. This may be traced to the fact that, at
$\alpha=\case12$, the energy density has the same scaling dimension at the
surface as in the bulk. In the energy sector the system does not feel anymore
the surface and behaves asymptotically as a periodic one. The
profile~\ref{e4.3} is recovered using~\ref{e2.7} with $x_{\rm e}^{\rm
s}=x_{\rm e}=1$. Actually, the numerical data display logarithmic corrections.
Good fits of the finite-size results were obtained with the functional form
$e(z)=A_0+A_1\ln(\sin z)$ where $A_0\gg A_1$. 
\item{\hglue-2mm(iii)}  For $\alpha>\case12$, i.e., in the first-order regime,
equation~\ref{e4.1} leads to the following expression for the energy density:
$$
e(z)=\frac{B}{L^{1/2+\alpha}}\left[\cos^{2\alpha+1}
\left(\frac{z}{2}\right)+\sin
^{2\alpha+1}\left(\frac{z}{2}\right)\right]\,.
\label{e4.4}
$$
Near the surface $e(z)$ scales as $L^{-(1/2+\alpha)}$, i.e., according
to~\ref{e3.18}, the gap-exponent relation is satisfied but the profile
has a functional form which clearly disagrees with the conformal expression
in~\ref{e2.7}. This is due to the localization near $z=\pi$ of the lowest
eigenstate $\psi_0(z)$, with vanishing eigenvalue in the continuum limit.
This anomalous behaviour is related to the persistence of surface order at
criticality. The next two eigenstates are extended and have a normal scaling
behaviour. Let us consider the corresponding off-diagonal
energy profile,
$$
e_{12}(z)=\psi_{1}(z)\psi_{2}(\pi-z)+\psi_{1}(\pi-z)\psi_{2}(z)\,.
\label{e4.5}
$$
Inserting~\ref{e3.19} with $k=1,2$ yields:
$$
e_{12}(z)=\frac{C}{L}(\sin z)^{2\alpha+1}\,.
\label{e4.6}
$$
This is to be compared with~\ref{e2.7} with $x_{\rm e}=1$ and the approriate
surface scaling dimension, $(x_{\rm e}^{\rm s})_{12}$, which is {\it not}
given by~\ref{e3.18} but can be deduced from the gap-exponent
relation~\ref{e2.8} and reads:
$$
(x_{\rm e}^{\rm s})_{12}=\frac{L}{\pi}(\Lambda_1+\Lambda_2)=2\alpha+2\,.
\label{e4.7}
$$
Thus we find an agreement with the behaviour expected from conformal
invariance,
if we leave out the anomalous first excitation. This point is discussed
further in the final section where a parallel is made with the analogous
situation for the homogeneous system with fixed boundary conditions.

The energy density profiles obtained
numerically in the two regimes using equation~\ref{e4.1} for
$\alpha<\case12$ and equations~\ref{e4.1} and~\ref{e4.5} for $\alpha>\case12$
are shown in figures~\ref{fig1}--\ref{fig3}, respectively.

\subsection{Magnetization profiles}
The off-diagonal magnetization profile is given by the matrix element
$m(l)=\langle 0\vert\sigma_l^{x}\vert\sigma\rangle$, where
$\vert\sigma\rangle=\eta_0^{\dagger}\vert 0\rangle$ is the lowest excited state
in the odd sector of $\cal H$ with energy $E_{\sigma}$.  Rewriting
$\sigma_l^{x}$ in terms of diagonal fermions and using Wick's theorem, the
local magnetization can be expressed as a determinant~\cite{berche96}
$$
m(l)=\left\vert\matrix{
H_{1}& G_{11}& G_{12}& \dots &G_{1l-1}\cr
H_{2}& G_{21}& G_{22}& \dots &G_{2l-1}\cr
\vdots& \vdots&\vdots& &\vdots \cr
H_l& G_{l1}& G_{l2}& \dots &G_{ll-1}\cr
}
\right\vert
\label{e4.8}
$$
with
$$
H_{j}=\phi_0(j)\qquad G_{ij}=-\sum_k\phi_k(i)\psi_k(j)
\label{e4.9}
$$
where the vectors ${\bphi}_k$ and ${\bpsi}_k$ are the eigenvectors defined
in~\ref{e3.7}. Since $\sigma_l^{x}$ is non-local in terms of the
diagonal fermion operators, $\eta_k^{\dagger}$ and $\eta_k$, the expression of
$m(l)$ is much more complicated than the corresponding
one~\ref{e4.1} for the energy density and we had to study the behaviour of
the magnetization profile numerically.

%%%%%%%% figure 4 %%%%%%%%%%%%%%%%%%%%%%%%%%%%%%%%%%%%%%%%%%%%%
{\par\begingroup\parindent=0pt\medskip
\epsfxsize=7.5truecm
\topinsert
\centerline{\epsfbox{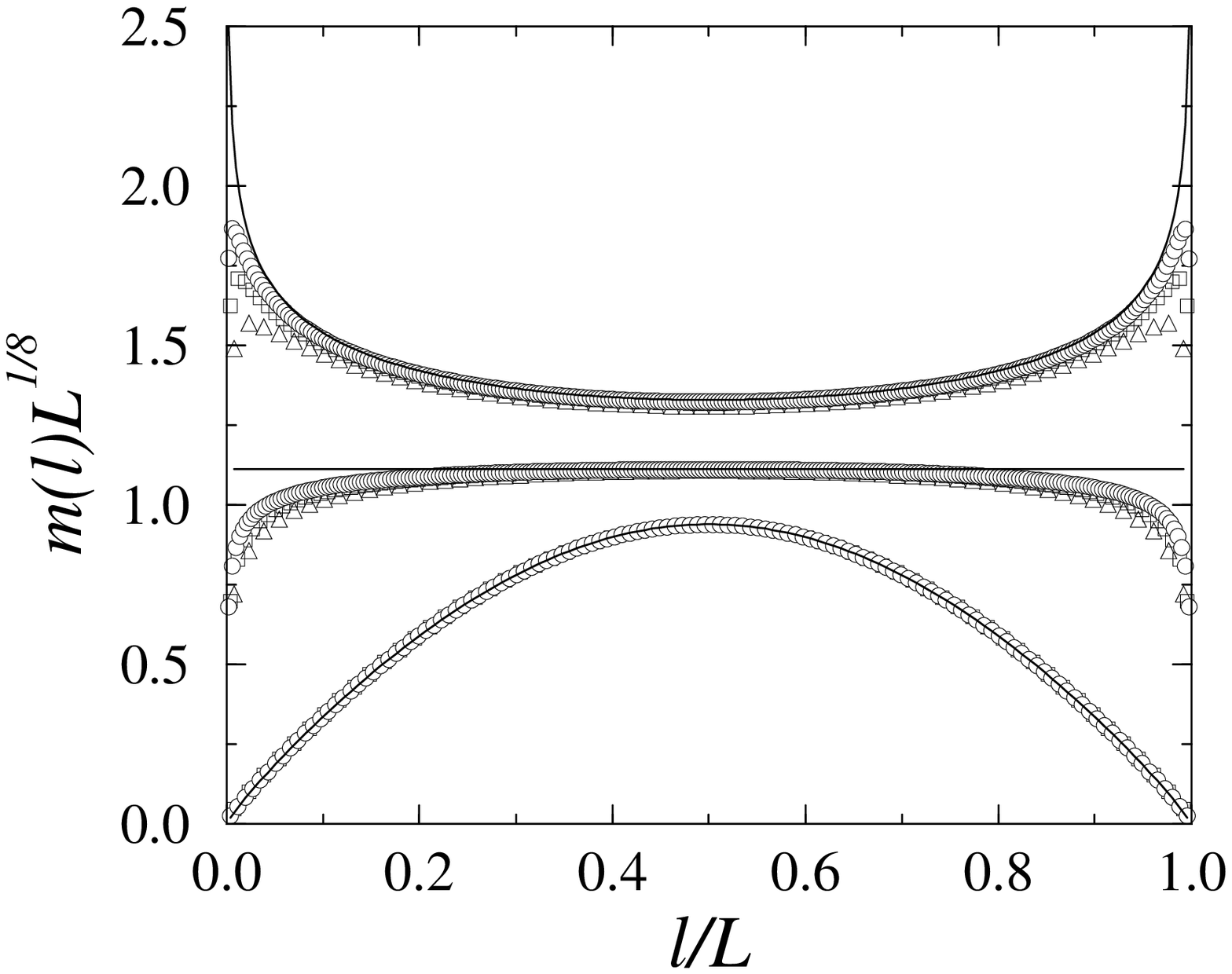}}
\smallskip
\figure{Rescaled magnetization profiles with $\alpha=\case32$,
$\alpha=\case38$ and $\alpha=-\case12$ from top to bottom
for chain sizes $2^6$ (triangle), $2^7$ (square) and $2^8$ (circle).
The full curves give the corresponding conformal profiles.
\label{fig4}}
\endinsert
\endgroup
\par}
%%%%%%%%%%%%%%%%%%%%%%%%%%%%%%%%%%%%%%%%%%%%%%%%%%%%%

The rescaled profiles shown in figure~\ref{fig4}, obtained for  different
values of $\alpha$,  converge with increasing sizes towards
the conformal expression which, according to \ref{e2.7} with $x_{\rm
m}=\case18$ for the bulk magnetic exponent, takes the form
$$
m(z)=\frac{A}{L^{1/8}}(\sin z)^{x_{\rm m}^{\rm s}-1/8}
\label{e4.10}
$$
where $z=\pi l/L$.

For $\alpha<\case12$, the critical exponent $x_{\rm m}^{\rm s}=\case12-\alpha$
has to be used. In particular, the value $\alpha=\case38$ corresponds to equal
surface and bulk scaling dimensions,
$x_{\rm m}^{\rm s}=x_{\rm m}=\case18$, and then~\ref{e4.10} leads to a
constant magnetization  profile.
The deviations near the surfaces in figure~\ref{fig4}, which are
slowly decreasing with increasing sizes, are due to finite-size corrections.

In the first-order regime, $\alpha>\case12$,
the critical profile has the same form as for the  homogeneous
system with fixed boundary conditions since $x_{\rm m}^{\rm s}=0$ in both
cases. The rescaled profile is divergent at the surface in the continuum
limit, the numerical data obtained on the discrete system display this
tendency with increasing sizes.

\section{Discussion}

The anomalous behaviour of the energy-density profile in the first-order
regime, $\alpha>\case12$, is linked to the localization, near $z=0$, of the
state $\psi_{0}(z)$. The corresponding excitation, $\Lambda_{0}$, vanishes as
$L^{-2\alpha}$, i.e., faster than higher excitations which display the usual
$L^{-1}$ scaling behaviour. The localization of the lowest eigenstate is
itself related to the persistence of surface order at criticality. This
suggests an analogy with the homogeneous system with symmetric fixed
boundary conditions for which, similarly, the lowest excitation vanishes and
corresponds to a mode which is completely localized at the surface.
The energy spectrum is doubly degenerate and the
first mode does not contribute to the conformal profile which is given by
the matrix element $\langle 0|\sigma^{z}|\epsilon_{++}\rangle$, where
$|\epsilon_{++}\rangle=\eta_{2}^{\dagger}\eta_{1}^{\dagger}|0\rangle$ is the first
non-trivial even state \cite{turban97}. The spectrum and eigenvectors are
those of the dual chain with free boundary conditions. In the continuum
limit, the energy-density profile is given by \ref{e2.7} with $x_{e}=1$ and
$(x_{\rm e}^{\rm s})_{++}=L/\pi(\Lambda_{1}+\Lambda_{2})=2$, i.e.,
$$
e_{++}(z)=\frac{A}{L}\sin z\,.
\label{e5.1}
$$
This conformal profile corresponds to $e_{12}(z)$ in the HvL model. The
two systems differ only by the values of the surface scaling dimensions for
the energy: to $(x_{\rm e}^{\rm s})_{++}=2$ in the model with fixed
boundary conditions corresponds $x_{\rm e}^{\rm s}=2+2\alpha$ in the HvL
model. The
difference comes from the shift between the two excitation spectra,
$\Lambda_{k}=(\Lambda_{k})_{++}+\alpha$. Since in both cases the
surface perturbation does not change the bulk critical behaviour, the bulk
exponents appearing in the profiles are those of the  homogeneous Ising
quantum chain.

The surface magnetic exponent, $x_{\rm m}^{\rm s}$, vanishes for the two
problems due to surface order. Thus, if conformal invariance holds as
indicated by the numerical data, one expects the two magnetization profiles to
differ only through the value of their amplitude, which is $\alpha$-dependent
in the HvL model and, asymptotically, we have
$$
m(l,\alpha)\sim m_{++}(l)=\left\vert\matrix{ 1& 0& 0& \dots &0\cr 0&
(G_{21})_{++}& (G_{22})_{++}& \dots &(G_{2l-1})_{++}\cr  \vdots&
\vdots&\vdots& &\vdots \cr 0& (G_{l1})_{++}& (G_{l2})_{++}& \dots
&(G_{ll-1})_{++}\cr } \right\vert
\label{e5.2}
$$
where $(G_{i,j})_{++}$ is defined as in~\ref{e4.9} and the
index $++$ refers to the homogeneous chain with fixed boundary conditions.
This expression has to be compared to~\ref{e4.8} where $\phi_{0}(1)=\Or(1)$
and  $\phi_{0}(n)=\Or(L^{-\alpha})$ when $z=n\pi/L=\Or(1)$. All the information
about the inhomogeneity is then contained in the  non-universal amplitude
$A(\alpha)$ which is increasing with $\alpha$, like the surface critical
magnetization.

To summarize, we have studied the critical energy and magnetization profiles
of the HvL model, using the methods of conformal invariance 
and through direct analytical and numerical calculations. We have shown that
the scaling form of the profiles, resulting from conformal invariance, remains
valid for the marginally perturbed system. In the regimes of second- and
first-order surface transitions, the profiles keep the same form as for the
unperturbed system with free and fixed boundary conditions, respectively,
with the HvL values for the surface exponents.

\ack  The Laboratoire de Physique des Mat\'eriaux is Unit\'e Mixte de
Recherche CNRS No 7556. The work of FI has been supported by the Hungarian
National Research Fund under grant Nos OTKA TO23642, TO25139, MO28418 and by
the Ministery of Education under grant No FKFP 0596/1999.

\numreferences
\bibitem{belavin84} {Belavin A A, Polyakov A M and Zamolodchikov A B 1984}
{\NP} B {\bf  241} 333
\bibitem{cardy84} {Cardy J L 1984} {\NP} B {\bf 240} 514
\bibitem{turban85} {Turban L 1985} {\JPA} {\bf 18} L325
\bibitem{henkel87} {Henkel M and Patk\'os A 1987} {\JPA} {\bf 20} 2199
\bibitem{burkhardt90} {Burkhardt T W and Igl\'oi F 1990} {\JPA} {\bf
23} L633
\bibitem{igloi90} {Igl\'oi F, Berche B and Turban L 1990} {\PRL} {\bf
65} {1773}
\bibitem{bariev91a} {Bariev R Z and Peschel I 1991} {\JPA} {\bf 24} L87
\bibitem{turban91} {Turban L 1991} {\PR} B {\bf 44} 7051
\bibitem{bariev91b} {Bariev R Z and Peschel I 1991} {\it Phys. Lett.} A
{\bf 153} 166
\bibitem{turban93} {Turban L and Berche B 1993} {\JPA} {\bf
26} 3131
\bibitem{hilhorst81} {Hilhorst H J and van Leeuwen J M J 1981}
{\PRL} {\bf 47} 1188
\bibitem{blote83} {Bl\"ote H W J and Hilhorst H J 1983} {\PRL} {\bf 51}
2015 
\bibitem{burkhardt84a} {Burkhardt T W and Guim I 1984} {\PR} B {\bf
29} 508
\bibitem{burkhardt84b} {Burkhardt T W, Guim I, Hilhorst H J and van
Leeuwen J M J 1984} {\PR} B {\bf 30} 1486
\bibitem{peschel84} {Peschel I 1984} {\PR} B {\bf 30} 6783
\bibitem{blote85} {Bl\"ote H W J and Hilhorst H J 1985} {\JPA} {\bf 18}
3039
\bibitem{bariev88} {Bariev R Z 1988} {\it Sov. Phys.--JETP} {\bf 67} 2170
\bibitem{igloi90} {Igl\'oi F 1990} {\PRL} {\bf 64} 3035
\bibitem{berche90} {Berche B and Turban L 1990} {\JPA} {\bf 23} 3029
\bibitem{peschel92} {Peschel I and Wunderling R 1992} {\AP} {\bf 1} 125
\bibitem{igloi93} {Igl\'oi F, Peschel I and Turban L 1993} {\it Adv.
Phys.} {\bf  42} 683
\bibitem{burkhardt82a} {Burkhardt T W 1982} {\PRL} {\bf 48} 216
\bibitem{cordery82} {Cordery R 1982} {\PRL} {\bf 48} 215
\bibitem{burkhardt82b} {Burkhardt T W 1982} {\PR} B {\bf 25} 7048
\bibitem{turban97} {Turban L and Igl\'oi F 1997} {\JPA} {\bf 30} L105
\bibitem{kogut79} {Kogut J B 1979} {\RMP} {\bf 51} 659
\bibitem{cardy83} {Cardy J L 1983} {\JPA} {\bf 17} L385
\bibitem{burkhardt85} {Burkhardt T W and Eisenriegler E 1985} {\JPA} {\bf 18}
L83 
\bibitem{barber84} {Barber M N, Peschel I and Pearce PA 1984} {\it J. Stat.
Phys.} {\bf 37} 497 
\bibitem{jordan28} {Jordan P and Wigner E 1928} {\ZP} {\bf
47} 631 \bibitem{lieb61} {Lieb E H, Schultz T D and Mattis D C 1961} {\APNY}
{\bf 16} 406
\bibitem{pfeuty70} {Pfeuty P 1970} {\it Ann. Phys. Paris} {\bf 57} 79
\bibitem{abramowitz70} {Abramowitz M and Stegun I A 1970} {\it Handbook of
Mathematical  Functions} 9th edn (New York: Dover)
\bibitem{berche96} {Berche P E, Berche B and Turban L 1996} {\it J. Phys. I}
{\bf 6} 621

\vfill\eject\bye